\documentclass[journal, onecolumn, draftcls]{IEEEtran}

\usepackage{amsfonts}
\usepackage{mathtools}
\usepackage{wasysym}

\usepackage{amssymb}
\usepackage{amsmath}
\usepackage{bbm}
\usepackage{bm}
\usepackage{dsfont}
\usepackage{graphicx}

\usepackage{enumitem}
\usepackage{tikz}
\usepackage{subfigure}
\usepackage{ifpdf}

\newcommand{\Z}{\mathbb{Z}}

\newcommand{\Code}{\bm{\mathcal{C}}}

\newcommand{\wH}{w_{\text{H}}}

\newcommand{\aleps}{\alpha_{\epsilon}}

\newcommand{\Esp}{E_{\text{sp}}}
\newcommand{\Eex}{E_{\text{ex}}}
\newcommand{\Ex}{E_{\text{x}}}

\newcommand{\Rcrit}{R_{\text{crit}}}
\newcommand{\Pe}{\mathsf{P}_{\text{e}}}

\newcommand{\Pemax}{\mathsf{P}_{\text{e,max}}}

\newtheorem{theorem}{Theorem}

\newtheorem{remark}{Remark}

\newcommand{\matc}{\ensuremath{\mathcal{C}}}

\newcommand{\PP}{\ensuremath{\mathbb{P}}}

\ifx\eqref\undefined
	\newcommand{\eqref}[1]{~(\ref{#1})}
\fi
\ifx\mod\undefined
	\def\mod{\mathop{\rm mod}}
\fi

\def\exp{\mathop{\rm exp}}

\def\EE{\mathbb{E}\,}

\def\PP{\mathbb{P}}

\def\eqdef{\stackrel{\triangle}{=}}

\ifx\lesssim\undefined
\def\simleq{{{\mskip 3mu plus 2mu minus 1mu%
	\setbox0=\hbox{$\mathchar"013C$}%
	\raise.2ex\copy0\kern-\wd0%
	\lower0.9ex\hbox{$\mathchar"0218$}}\mskip 3mu plus 2mu minus 1mu}}
\else
\def\simleq{\lesssim}
\fi
\ifx\gtrsim\undefined
\def\simgeq{{{\mskip 3mu plus 2mu minus 1mu%
	\setbox0=\hbox{$\mathchar"013E$}%
	\raise.2ex\copy0\kern-\wd0%
	\lower0.9ex\hbox{$\mathchar"0218$}}\mskip 3mu plus 2mu minus 1mu}}
\else
\def\simgeq{\gtrsim}
\fi

\newif\ifmapx
{\catcode`/=0 \catcode`\\=12/gdef/mkillslash\#1{#1}}
\edef\jobnametmp{\expandafter\string\csname typewriter_journal_apx\endcsname}
\edef\jobnameapx{\expandafter\mkillslash\jobnametmp}
\edef\jobnameexpand{\jobname}
\ifx\jobnameexpand\jobnameapx
\mapxtrue
\else
\mapxfalse
\fi
\long\def\apxonly#1{\ifmapx{\color{blue}#1}\fi}

\begin{document}

\IEEEoverridecommandlockouts

%

\title{Bounds on the Reliability Function of Typewriter Channels}

\author{Marco~Dalai,~\IEEEmembership{Senior Member,~IEEE}, Yury~Polyanskiy~\IEEEmembership{Senior Member,~IEEE},
\thanks{M. Dalai is with the Department
of Information Engineering, University of Brescia, Italy, email: marco.dalai@unibs.it.
Y. Polyanskiy is with the Department of Electrical Engineering and Computer Science, Massachusetts Institute of
Technology, Cambridge, MA, USA, email: yp@mit.edu. This work was supported (in part) by the 
Center for Science of Information (CSoI), an NSF Science and Technology Center, under grant agreement CCF-09-39370, by the NSF grant CCF-13-18620 and by the Italian Ministry of Education under grant PRIN 2015 D72F16000790001.
Part of the results were first presented in \cite{dalai-polyanskiy-2016}.
}}

\maketitle

\begin{abstract}

New lower and upper bounds on the reliability function of typewriter channels are given.  Our lower bounds improve upon
the (multiletter) expurgated bound of Gallager, furnishing a new and simple counterexample to a conjecture made in 1967
by Shannon, Gallager and Berlekamp on its tightness.  The only other known counterexample is due to Katsman, Tsfasman
and Vl\u{a}du\c{t} who used algebraic-geometric codes on a $q$-ary symmetric channels, $q\geq 49$. Here we prove, by
introducing dependence between codewords of a random ensemble, that the conjecture is false even for a typewriter
channel with $q=4$ inputs. In the process, we also demonstrate that Lov\'asz's proof of the capacity of the pentagon 
was implicitly contained (but unnoticed!) in the works of Jelinek and Gallager on the
expurgated bound done at least ten years before Lov\'asz.  In the opposite direction, new upper
bounds on the reliability function are derived for channels with an odd number of inputs by using an adaptation of
Delsarte's linear programming bound. First we derive a bound based on the minimum distance, which combines Lov\'asz's
construction for bounding the graph capacity with the McEliece-Rodemich-Rumsey-Welch construction for bounding the
minimum distance of codes in the Hamming space. Then, for the particular case of cross-over probability $1/2$, we derive
an improved bound by also using the method of Kalai and Linial to study the spectrum distribution of codes.
\end{abstract}

\section{Introduction}

Consider the typewriter channel $W$ whose input and output alphabets are $\mathbb{Z}_q$, and whose transition probabilities are 
\begin{equation}\label{eq:typewriter}
W(y|x)=
\begin{cases}
1-\epsilon & y=x\\
\epsilon & y=x+1 \mod q
\end{cases}
\end{equation}
where, without loss of generality we assume through the paper that $0<\epsilon\leq 1/2$. We also assume $q\geq 4$, for reasons which will be clear in what follows.

This paper deals with the study, for these particular channels, of the classic problem of bounding the reliability function $E(R)$, defined by (\cite{shannon-gallager-berlekamp-1967-1}, \cite{gallager-book})
\begin{equation*}
E(R)=\limsup_{n\to\infty} \frac{1}{n}\log \frac{1}{\Pe(\lceil 2^{n R} \rceil, n)}\,,
\end{equation*}
where $\Pe(M, n)$  is the smallest possible probability of error of codes with $M$ codewords of length $n$. In particular, since the definition of $E(R)$ does not depend on whether one considers maximal or average probability of error over codewords (see \cite{shannon-gallager-berlekamp-1967-1}), we will use one quantity or the other according to convenience. In this paper all logarithms are to the base 2 and rates are thus measured in bits per channel use.

Bounding $E(R)$ for the considered channels needs first a discussion of their capacity and zero-error capacity. For any $q$, the capacity of the channel has the simple expression $C=\log(q)-H(\epsilon)$, where $H$ is the binary entropy function. Furthermore, for $q\geq 4$, those channels have a positive zero-error capacity $C_0$ \cite{shannon-1956}, which is defined as the highest rate at which  communication  is possible with probability of error precisely equal to zero. For even $q$, it is easily proved that $C_0=\log(q/2)$, while for odd $q\geq 5$ determining $C_0$ is a much harder problem. For $q=5$ Shannon \cite{shannon-1956} gave the lower bound $C_0\geq \log\sqrt{5}$, which Lov\'asz proved to be tight more than twenty years later \cite{lovasz-1979}. For larger odd values of $q$, Shannon observed that standard information theoretic arguments imply $C_0\leq \log(q/2)$, while Lov\'asz \cite{lovasz-1979} gave a better upper bound of the form $C_0\leq \log \theta(C_q)$, where $\theta(G)$ is the Lov\'asz theta function of a graph $G$ and $C_q$ is the cycle of length $q$, for which
\begin{equation}
\theta(C_q)=\frac{\cos(\pi q)}{1+\cos(\pi q)}q.
\label{eq:lovasz_theta_cq}
\end{equation}
Good lower bounds on $C_0$ for odd values of $q$ are also difficult to derive. Specific results have been obtained for example in \cite{baumert-et-al-1971}, \cite{bohman-2003-I}, \cite{mathew-ostergard-2016}, but there does not seem to be a sufficiently general result which singles out as the best for all odd $q$.

The focus of this paper is on the discussion of known bounds on $E(R)$ and on the derivation of new lower and upper
bounds. Specifically, the paper is structured as follows. In Section \ref{sec:class_bd} we discuss the classical upper
and lower bounds on the reliability function $E(R)$. Evaluation of the expurgated bound is non-trivial and requires 
deducing some observations which seemingly have not appeared in the literature.
In particular it is observed that the zero-error capacity of the pentagon can be
determined by a careful study of the expurgated bound, something which could have been done at least ten years before
Lov\'asz's paper settled the question. Then, in Section \ref{sec:new_LB} we present an improved lower bound for the case of even
$q$, showing that it also is a precisely shifted version of the expurgated bound for the BSC. The technique also applies
in principle to odd values of $q$ and we show in particular the result obtained for $q=5$. This result also provides an
elementeary disproof of the conjecture suggested in \cite{shannon-gallager-berlekamp-1967-1} that the expurgated bound
might be asymptotically tight when computed on arbitrarily large blocks, a conjecture which had been already disproved
in \cite{katsman-tsfasman-vladut-1992} by means of algebraic geometric codes.

In Section \ref{sec:UPs} we discuss upper bounds. Section~\ref{sec:UPs_binary} shows an error-exponent bound by 
extracting a binary subcode. Then in Section \ref{sec:new_UP_distance} we present a new upper bound for the case of odd $q$
based on the minimum distance of codes.  We use Delsarte's linear programming method \cite{delsarte-1973}
combining the construction used by Lov\'asz \cite{lovasz-1979} for bounding the graph
capacity with the construction used by McEliece-Rodemich-Rumsey-Welch \cite{mceliece-et-al-1977} for bounding the
minimum distance of codes in Hamming spaces. Finally, in Section \ref{sec:new_UP_spectrum} we give an improved upper
bound for the case of odd $q$ and $\epsilon=1/2$ following ideas of Litsyn~\cite{litsyn-1999}, see also
Barg-McGregor~\cite{barg-mcgregor-2005}, which in turn are based on estimates for the spectra of codes originated in
Kalai-Linial~\cite{kalai-linial-1995}.

\section{Classical bounds and Shannon-Gallager-Berlekamp conjecture}\label{sec:class_bd}

\subsection{Background on random coding bounds}
In~\cite{gallager-1965} Gallager showed that for an arbitrary DMC $W(y|x)$ there exists a blocklength-$n$ code of rate $R$ with
average probability of error bounded by
$$ P_e \le \exp\left\{-n E_{\text{r}}\left(R\right)\right\}\,, $$
where 
\begin{align}
E_{\text{r}}(R) & = \max_{0\leq \rho \leq 1 } E_0(\rho)-\rho R \label{eq:er_std}\\
	E_0(\rho) & = \max_{P}  \left[ -\log \sum_y \left(\sum_x P(x) W(y|x)^{1/{(1+\rho)}}\right)^{1+\rho} \right] \,.
\end{align}

For low rates Gallager also proved an improved (expurgated) bound given by:
\begin{equation}\label{eq:gall_expur}
	P_e \le \exp\left\{-n \Eex^k\left(R - {\log 4\over n}\right)\right\}\,, 
\end{equation}
for any $n$ which is a multiple of $k\ge 1$, where $k$ is an arbitrary positive integer and 
\begin{align}
\Eex^k(R)&= \sup_{\rho\geq 1} \Ex^k(\rho)-\rho R\label{eq:def-Eex^n} \\
\Ex^k(\rho) & =-\frac{\rho}{k}\log\min_{P_{X^k}}Q^k(\rho,P_{X^k})\\
Q^k(\rho,P_{X^k}) & =\sum_{\bm{x}_1,\bm{x}_2}
 P_{X^k}(\bm{x}_1)P_{X^k}(\bm{x}_2)g_k(\bm{x}_1,\bm{x}_2)^{1\over\rho}\\
g_k(\bm{x}_1,\bm{x}_2) &=\sum_{\bm{y}}\sqrt{\bm{W}(\bm{y}|\bm{x}_1)\bm{W}(\bm{y}|\bm{x}_2)},\label{eq:defgn}
\end{align}
and where $\bm{W}$ is the $k$-fold memoryless extension of $W$. This results in the following lower bound on the
reliability function:
\begin{align} E(R) &\ge  \Eex^\infty(R)\\
\Eex^\infty(R) &\eqdef \sup_{k \in \mathbb{Z}_+} \Eex^k(R) = \lim_{k\to\infty} \Eex^k(R) \label{eq:ex_inf}\,,
\end{align}
where the equality in~\eqref{eq:ex_inf} follows from
super-additivity of $k\Eex^k(R)$ and Fekete's lemma.\footnote{Super-additivity follows from taking $P_{X^{n+m}} =P^{(1)}_{X^n} \times P^{(2)}_{X^m}$, where
$ P^{(i)}$ are optimal inputs for lengths $n$ and $m$, respectively.} For a general channel computing $\Eex^\infty$ is
impossible due to maximization over all $k$-letter distributions, and hence most commonly this bound is used in the
weakened form by replacing $\Eex^\infty$ with $\Eex^1$.\footnote{Note that the exponent $E_{\text{r}}$ in~\eqref{eq:er_std} does
not change if we propose a similar $k$-letter extension: the optimal distribution $P_{X^k}$ may always be chosen to be a
product of single-letter ones.}

For understanding our results it is important to elaborate on Gallager's proof of~\eqref{eq:gall_expur}. 
Consider an arbitrary blocklength-$n$ 
code with $M$ codewords $\bm{x}_1,\ldots,\bm{x}_M$ and maximum-likelihood decoder $\hat m:\mathcal{Y}^n\to\{1,2,\ldots,M\}$. Define
$$ P_{i|j} \eqdef \PP[\hat m(Y^n)=i | X^n = \bm{x}_j]$$
to be the probability of detecting codeword $i$ when $j$ was sent. A standard upper bound on this probability
\cite[(5.3.4)]{gallager-book} is given by 
\begin{equation}
P_{i|j} \leq g_n(\bm{x}_i,\bm{x}_j)\,,\label{eq:pairwise}
\end{equation}
where $g_n$ was defined in~\eqref{eq:defgn}.
For the typewriter channel~\eqref{eq:typewriter} we can express the pairwise bound~\eqref{eq:pairwise} equivalently as
\begin{equation}
P_{i|j}\leq \aleps^{d(\bm{x}_i,\bm{x}_j)}\,,\label{eq:bhatta}
\end{equation}
where 
\begin{equation}
\aleps\eqdef \sqrt{\epsilon(1-\epsilon)}\,,
\end{equation} we agree that $\aleps^{\infty}=0$, and $d:\mathbb{Z}_q\times
\mathbb{Z}_q\to\{0, 1,\infty\}$ is a semidistance defined as
\begin{equation*}
d(x_1,x_2)\eqdef
\begin{cases}
0 & x_1=x_2\\
1 & x_1-x_2=\pm 1\\
\infty & x_1-x_2\neq \pm 1
 \end{cases}
\end{equation*}
and extended additively to sequences in $\mathbb{Z}_q^n$ 
\begin{equation*}
d(\bm{x}_1,\bm{x}_2)\eqdef\sum_k d(x_{1,k},x_{2,k}).
\end{equation*}

The average probability of error of the code can then be bounded using the union bound as
\begin{align}
\Pe & \leq \frac{1}{M}\sum_i \sum_{j\neq i} \aleps^{d(\bm{x}_i,\bm{x}_j)}\\
& =  \sum_{z=0}^n A_z \aleps^{z}\,,
\label{eq:Pefromspectrum}
\end{align}
where $A_z$ is the spectrum of the code
\begin{equation*}
A_z=\frac{1}{M}\left|\{(i,j):  i\neq j\,, d(\bm{x}_i,\bm{x}_j)=z\} \right|.
\end{equation*}

From expression~\eqref{eq:Pefromspectrum} one may get existence results for good codes by (for example), selecting
$\bm{x}_i$ randomly according to some i.i.d. distribution $(P_X)^n$ and averaging~\eqref{eq:Pefromspectrum}.
Gallager~\cite{gallager-1965} observed that for low rates the dominant term in the summation may correspond to $z$ such
that $\EE[A_z] \ll 1$. By expurgating from the code all pairs of codewords at distances $z$ s.t.
$\EE[A_z] \ll 1$ he obtained the exponential improvement~\eqref{eq:gall_expur}.

\begin{remark}[Shannon-Gallager-Berlekamp conjecture]
In~\cite{shannon-gallager-berlekamp-1967-1} it was conjectured that the hard-to-evaluate quantity $\Eex^\infty(R)$
equals the true reliability function for rates below the critical one.\footnote{Quoting from
\cite{shannon-gallager-berlekamp-1967-1}: ``The authors would all tend to conjecture [...] As yet there is little
concrete evidence for this conjecture.''} For symmetric channels it would be implied by the (conjectured) tightness of the
Gilbert-Varshamov bound. The conjecture was disproved by Katsman, Tsfasman and Vl\u{a}du\c{t}
\cite{katsman-tsfasman-vladut-1992} using algebraic-geometric codes which also beat the Gilbert-Varshamov bound (for alphabets with $49$ or more symbols). To the best
of our knowledge, no other disproof is known in the literature. The bound we provide in the next Section proves in
particular that $E(R)>\Eex^\infty(R)$ in some rate range for all typewriter channels for which we could compute
$\Eex^\infty(R)$ exactly (among which $q=4,5$), and hence it offers a second disproof of the conjecture.  The main
innovation of our approach is that our ensemble of codewords $\{\bm{x}_1,\ldots,\bm{x}_M\}$ has carefully designed
dependence between codewords. Otherwise, we do still rely on~\eqref{eq:Pefromspectrum}.
\end{remark}

\subsection{Evaluating classical bounds for the typewriter channel}

To calculate the random-coding exponent $E_{\text{r}}(R)$ one needs to notice that due to the symmetry of the typewriter
channel~\eqref{eq:typewriter} the optimal input distribution is uniform. In this way we get in parametric form over $\rho
\in [0,1]$, cf.~\cite[(46)-(50)]{gallager-1965}:
\begin{align}
E_{\text{r}}(R)& = \begin{cases} 
		\log\left( \frac{q}{1+2\sqrt{\epsilon(1-\epsilon)}}\right)-R\,, & R \le \Rcrit\\
		D(\epsilon_\rho\|\epsilon), \qquad & \Rcrit \le R=R_\rho \le C\\
	\end{cases}\label{eq:Er_typewriter_form}\\
\epsilon_\rho &  = \frac{\epsilon^{1/(1+\rho)}}{\epsilon^{1/(1+\rho)}+(1-\epsilon)^{1/(1+\rho)}}\label{eq:epsrho}\\
R_\rho & =\log(q)-h_2(\epsilon_\rho)\label{eq:Rrho}\\
\Rcrit &= \log(q)-h_2\left(\frac{\sqrt{\epsilon}}{\sqrt{\epsilon}+\sqrt{1-\epsilon}}\right)\\
C &= \log(q)-h_2(\epsilon)\\
h_q(x) & \eqdef x \log(q-1) - x \log x - (1-x)\log(1-x)\,.
\end{align}
One should notice that $E_{\text{r}}(R)$ coincides with the random coding exponent for the BSC just shifted by $\log(q/2)$ on the $R$
axis (extending of course the straight line portion at low rates down to $R=0$). 

A general upper-bound on $E(R)$ is the so-called sphere-packing bound~\cite{shannon-gallager-berlekamp-1967-1}, which for
the typewriter channel can be computed in a similar parametric form over $\rho \in [0,\infty)$:
\begin{equation}\label{eq:esp_tw}
	\Esp(R) = \begin{cases}
	\infty, & R < \log {q\over2}\\
	D(\epsilon_\rho\|\epsilon)\,, & \log{q\over2} \le R=R_\rho \le C\\
	\end{cases}\,,
\end{equation}
where $R_\rho$ and $\epsilon_\rho$ are as given by~\eqref{eq:epsrho} and~\eqref{eq:Rrho}. Note that again, $\Esp(R)$
corresponds to the sphere packing bound for the BSC shifted by a $\log(q/2)$ quantity on the rate axis.  

We proceed to evaluating the expurgated bound. 
As we mentioned above, evaluating $\Eex^\infty(R)$ is generally non-trivial due to the necessity of optimizing
over multi-letter distributions. We need, therefore, to use the special structure of the channel. 
For the BSC with parameter $\epsilon$, for example, Jelinek~\cite{jelinek-1968} proved that $\Eex^n(R)$ does not 
depend on $n$ and takes the form
\begin{equation}
\Eex^{\text{BSC}}(R)=
\begin{cases}
\log\frac{2}{1+2\aleps}-R ,&  R\geq R_{\text{ex}}^{*}(\epsilon;2)\\
-\delta_{GV}(R;2)\log(2 \aleps) ,&   R\leq R_{\text{ex}}^{*}(\epsilon;2)\,.
\end{cases}
\label{eq:Eex_BSC}
\end{equation}
where 
\begin{equation}
R_{\text{ex}}^{*}(\epsilon;2)=\log 2 - h_2\left(\frac{2 \aleps}{1+2\aleps}\right)\,,
\end{equation}
and $\delta_{GV}(R;q)$ is the Gilbert-Varshamov bound for $q$-ary codes defined by the condition
\begin{align} 
	R & = \log q - h_q(\delta_{GV}(R;q))\,.
\end{align}  

We next present our result on finding $\Eex^\infty(R)$ for typewriter channels.

\begin{theorem}\label{th:exinf} Let $\theta = q/2$ if $q$ is even and $\theta = q/(1+\cos(\pi/q)^{-1})$ if $q$ is odd.
Define $\bar \rho = {\log {\aleps} \over \log {q-\theta\over 2\theta}}$, then 
\begin{align} \Ex^n(\rho) &= \rho \log\left(\frac{q}{1+2\aleps^{1/\rho}} \right) && \text{if~} \rho \le \bar \rho
	\label{eq:exinf_1}\\
   \Ex^n(\rho) &=  \rho \log \theta\,, && \text{if~} \rho > \bar\rho \text{~and~$q$ even}\label{eq:exinf_2}\\
   \Ex^n(\rho) &\le \rho \log \theta\,, && \text{if~} \rho > \bar\rho \text{~and~$q$ odd}\label{eq:exinf_3}\\
   \Ex^n(\rho) &= \rho \log \theta\,, && \text{if~} \rho > \bar\rho, q=5 \text{~and $n$ even}
\end{align}   
\end{theorem}
\begin{remark} In short, for even $q$ the expurgated bound ``single-letterizes'', while for $q=5$ the asymptotics 
is already achieved at $n=2$. Note that for $q=5$ we do not compute $\Ex^n(\rho)$ for odd values of $n$, but 
due to super-additivity of $n\Ex^n$ and $n\Eex^n$ we may compute the limit along the
subsequence of even $n$. The $\theta$ we defined is precisely the Lov\'asz $\theta$-function for the $q$-cycle (a graph with $q$ vertices and $q$ edges connected to form a polygon with $q$ sides). How can a
$\theta$-function appear in the study of the expurgated bound, when the latter predates the former by a decade? 
See Remark~\ref{rem:on_expurgated_lovasz} below.
\end{remark}

\begin{remark}
Converting Theorem~\ref{th:exinf} to statements about $\Eex^n(R)$ is done via~\eqref{eq:def-Eex^n} and tedious algebra. 
Let $\bar\epsilon$ be the smallest $\epsilon$ for which
\begin{equation}
	\log\frac{q}{1+2\aleps}+\frac{2\aleps}{1+2\aleps}\log\aleps\leq \log\theta
\label{eq:Eex_condbareps}
\end{equation}
If $\epsilon \in [\bar\epsilon, 1/2]$, \textit{for even $q$} and any $n\geq 1$
\begin{equation}
 \Eex^n(R) = \Eex^\infty(R) = \begin{cases}
	\infty & R< \log(\theta)\\
	\log\frac{q}{1+2\aleps}-R  & R\geq \log(\theta)
	\end{cases}
	\label{eq:Eexeven_str}\,. 
\end{equation}
Furthermore, for $q=5$ the second equality holds and the first one holds for even $n$.
If $\epsilon < \bar\epsilon$ the expression above holds for rates \underline{outside} of the interval $\log\theta \leq R \leq R_{\text{ex}}^*(\epsilon;q)$, where
\begin{equation}
R_{\text{ex}}^*(\epsilon;q)= \log \frac{q}{1+2\aleps}+\frac{2\aleps}{1+2\aleps} \log \aleps\,, \qquad  \epsilon < \bar\epsilon\,,q\geq 4\,.
\label{eq:Eex_interval}
\end{equation}
Inside the interval $\log\theta \leq R \leq R_{\text{ex}}^*(\epsilon;q)$ we have (again with the same specifications on $q$ and $n$) the parametric representation 
\begin{align}
R & = \log\left(\frac{q}{1+2\aleps^{1/\rho}}\right)+\frac{2\aleps^{1/\rho}}{\rho(1+2\aleps^{1/\rho})}\log(\aleps)\\
\Eex^n(R)  &=\Eex^\infty(R) 	=\frac{2\aleps^{1/\rho}}{(1+2\aleps^{1/\rho})}\log\frac{1}{\aleps},
\end{align}
where $\rho$ runs in the interval $[1,\rho_0]$, $\rho_0$ being such that $\aleps^{1/\rho_0}=\alpha_{\bar{\epsilon}}$. Note that for even $q$, since $\theta=q/2$, $\bar{\epsilon}$ does not depend on $q$ (in particular, $\bar{\epsilon}\approx 0.022$); the functions $\Eex^n(R)$ all have the same shape and simply shift on the $R$ axis by 1 (bit) as $q$ moves from one even value to the next.
\textit{For all odd $q$}, the above expressions provide an upper bound on $\Eex^n(R)$ for all $n$ (which, again, is tight for $q=5$ and even $n$).
\apxonly{Here is more: 
Using the known properties of the expurgated bound \cite{gallager-1965} (or with just some tedious computations), we deduce that in the interval $[1,\bar{\rho}]$ the function $\Ex^n(\rho)$ is strictly concave and has derivative smaller than $\log \theta$ already at $\rho=1$ for $\epsilon> \bar\epsilon$. So, for $\epsilon\geq \bar{\epsilon}$ the supremum over $\rho$ in \eqref{eq:def-Eex^n} is achieved at $\rho=1$ for $R\geq \log\theta$ while it is infinite, and approached as $\rho\to\infty$, for $R<\log\theta$.
This proves the first part of the proposition. Explicit computation of the optimizing $\rho$ leads to the stated bound
for the case $\epsilon<\bar{\epsilon}$.}
\end{remark}

\begin{IEEEproof}[Proof of Theorem \ref{th:exinf}]
The idea is to use Jelinek's criterion~\cite{jelinek-1968} for single-letter optimality. 
Consider first the minimization of the quadratic form $Q^n(\rho,P_{X^n})$
and note that the $q^n\times q^n$ matrix with elements $g_n(\bm{x}_1,\bm{x}_2)^{{1/\rho}}$, call it $g_n^{\odot {1/\rho}}$,  is the $n$-fold Kronecker power of the $q\times q$ matrix
\begin{equation}
g_1^{\odot {1/\rho}}=\left(
\begin{array}{cccccc}
1 & \aleps^{1/\rho} & 0 & \cdots & 0  & \aleps^{1/\rho}\\
\aleps^{1/\rho} & 1 & \aleps^{1/\rho} & \cdots & 0  & 0\\
\vdots & \vdots & \vdots& \ddots & \vdots & \vdots\\
\aleps^{1/\rho} & 0 & \cdots & \cdots & \aleps^{1/\rho} & 1
\end{array}
\right).\label{eq:g1_1/rho}
\end{equation}
Note that if $g_1^{\odot 1/\rho}$ is a positive semidefinite matrix, so is $g_n^{\odot
{1/\rho}}$. In that case, the quadratic form defining $Q^n(\rho,P_{X^n})$ for any $n$ is a convex function of $P_{X^n}$. Jelinek \cite{jelinek-1968}
showed that it is minimized by a product distribution $P_{X^n}=P\times P \cdots\times P$, where $P$ is optimal for $n=1$,
and the achieved minimum is just the $n$-th power of the minimum achieved for $n=1$. Thus, if the matrix with elements
$g_1(x_1,x_2)^{1/\rho}$ is positive semidefinite, then $\Ex^n(\rho)=\Ex^1(\rho)$. Furthermore, in this case the
convexity of the quadratic form and the fact that  $g_1^{\odot {1/\rho}}$ is circulant imply that the uniform
distribution is optimal. Hence, by direct computation,
\begin{equation}\label{eq:Ex_pos_semidef}
\Ex^n(\rho)=\rho \log\left(\frac{q}{1+2\aleps^{1/\rho}} \right)
\end{equation}
whenever $g_1^{\odot 1/\rho}$ is positive semidefinite. The eigenvalues of $g_1^{\odot 1/\rho}$ are $\lambda_k=1+2
\aleps^{1/\rho}\cos(2\pi k/q)$, $k=0,\ldots, q-1$, and the matrix is positive semidefinite whenever $\rho \le \bar\rho$,
which proves~\eqref{eq:exinf_1}.

We now proceed to studying the case $\rho >\bar\rho$. First, note that 
$\Ex^n(\bar{\rho}) = \bar{\rho}\log\theta$. When $\rho$ exceeds $\bar{\rho}$ the matrix
$g_1^{\odot 1/\rho}$ has negative eigenvalues and the previous method of evaluation of $\Ex^n(\rho)$ does not apply. 
Instead, we observe that the minimum of
$Q^n(\rho,P_{X^n})$ is non-decreasing in $\rho$, and hence for $\rho>\bar{\rho}$
\begin{align}
\Ex^n(\rho) & \leq -\frac{\rho}{n}\log \min_{P_{X^n}}Q^n(\bar{\rho},P_{X^n})\label{eq:upperb_Exrho}\\
& = \frac{\rho}{\bar{\rho}}\Ex^n(\bar{\rho})\label{eq:upperb_Exrho2}\\
& = \rho \log \theta\,,\label{eq:upperb_Exrho3}
\end{align}
where the last step is obtained using the definition of $\bar{\rho}$ and equation \eqref{eq:Ex_pos_semidef}. This
establishes~\eqref{eq:exinf_3} and part of~\eqref{eq:exinf_2}. 

To show the equality in~\eqref{eq:exinf_2}, we assume $q$ is even and combine~\eqref{eq:upperb_Exrho3} with 
evaluation of the function $Q^1(\rho,P)$ when $P$ is the uniform distribution over the set $\{0,2,\ldots,q-2\}$. 
This results in $\Ex^n(\rho)\geq \Ex^1(\rho)=\rho \log(q/2)=\rho \log\theta$.

Finally, for the $q=5$ we pair~\eqref{eq:exinf_3} with evaluation of $Q^2(\rho,P_{X^2})$ by setting $P_{X^2}$ to be 
the uniform distribution on Shannon's zero-error code $\{(0,0), (1,2), (2,4), (3,1), (4,3)\}$. This is easily seen to
achieve equality in~\eqref{eq:upperb_Exrho3}.
\end{IEEEproof}

\begin{remark} [How Gallager missed discovering Lov\'asz's $\theta$-function]
\label{rem:on_expurgated_lovasz}
Let us denote $R_{x,\infty}^{(n)} = \sup\{R: \Eex^{n}(R)=\infty\}$, $R_{x,\infty}^* = \lim_n R_{x,\infty}^{(n)} = \sup\{R:
\Eex^{\infty}(R) = \infty\}$. Also let $C_{0,n}$ be the largest rate of a zero-error
code of blocklength $n$ and $C_0$ be the zero-error capacity of the channel (which is also the Shannon capacity of a
confusability graph). By taking $P_{X^n}$ to be uniform on the zero-error code, Gallager~\cite{gallager-1965} observed
already in 1965 that $R_{x,\infty}^{(n)} \ge C_{0,n}$ and thus $C_0 \le R_{x,\infty}^*$. 
Since Theorem \ref{th:exinf} finds $\Eex^\infty$ and hence $R_{x,\infty}^*$, it also shows that the Shannon capacity
of the pentagon is $\sqrt{5}$. In particular, we point out that this result is obtained by only using tools which were
already available in the '60s, at least ten years before Lov\'asz's paper \cite{lovasz-1979} appeared (!). Similarly,
Theorem \ref{th:exinf} implies the upper bound $C_0\leq \log(\theta)$ for $q$-cycles, 
where $\theta$ is precisely the Lov\'asz theta function. The reader might compare the statement of Theorem \ref{th:exinf} with the results in \cite[Sec. V.C]{dalai-TIT-2013}; for example, it implies the bound in \cite[page 8038, last equation]{dalai-TIT-2013} and it shows that \cite[eq. (27)]{dalai-TIT-2013} holds for a typewriter channel with even number of inputs even though the matrix $g_1^{\odot 1/\rho}$ is not positive semidefinite for all $\rho\geq 1$.

To complete the discussion, we mention that Gallager's bound implies also $C_{0,n} \ge R_{x,\infty}^{(n)} - {\log 4\over n}$
and thus $R_{x,\infty}^* = C_0$. In fact, we also have $R_{x,\infty}^{(n)} = C_{0,n}$, as shown by Korn~\cite{korn-1968}. 
 In combinatorics, this fact was discovered slightly earlier by  Motzkin and Strauss~\cite{motzkin1965maxima}. \apxonly{Precisely they have
shown that for any graph
$$ \max \{\sum_{i \sim j} x_i x_j: x_i \ge 0, \sum x_i = 1\} = {1\over 2} \left(1-{1\over \omega(G)}\right)\,.$$
}
\end{remark}

\begin{remark} Slightly generalizing the reasoning of Gallager and Jelinek, leads the following bound on the Shannon capacity of an arbitrary graph $G$:
	$$ C_0(G) \le \inf_M \sup_P \left(\sum_{i,j \in V(G)} P_i M_{i,j} P_j \right)^{-1}\,,$$
	where supremum is over all probability distributions on $V(G)$ and infimum is over all positive-semidefinite
	matrices $M$ with unit diagonal and $M_{i,j}=0$ whenever $i\neq j$ and $(i,j)\not\in E(G)$. This bound, in turn,
	is known to be equivalent to Lov\'asz's bound, see~\cite[Theorem 3]{mceliece-et-al-1978}. \apxonly{The
	restriction to $P_i\ge 0$ does not matter since we can always flip sign of rows/columns of $M_{i,j}$ to make
	sure optimizing $P_i$ even without positivity restriction is in fact positive.}
\end{remark}

\section{New lower (achievability) bound on $E(R)$}
\label{sec:new_LB}
In this section we provide new lower bounds on $E(R)$ for some typewriter channels.
Our new bounds are based on the idea of building codes which are the union of cosets of good zero-error codes. In particular, we improve Gallager's expurgated bound in all those cases in which we can evaluate $\Eex^\infty(R)$ exactly, namely when $q$ is even or $q=5$.

\begin{theorem}
Let $q$ be even. Then, for $R>\log(q/2)$ we have the bound
\begin{equation}
E(R)\geq \Eex^{\text{BSC}}(R-\log(q/2))\,,
\end{equation} 
where $\Eex^{\text{BSC}}(R)$ is the expurgated bound of the binary symmetric channel given in \eqref{eq:Eex_BSC}.
\label{th:GV-expurgatedbound}
\end{theorem}

\begin{IEEEproof}
We upper bound the error probability for a code by using a standard union bound on the probability of confusion among
single pairs of codewords~\eqref{eq:Pefromspectrum}. The code is built using a Gilbert-Varshamov-like procedure, though we exploit carefully the
properties of the channel to add some structure to the random code (i.e. we introduce dependence among codewords) 
and obtain better results than just picking random independent codewords. 

A code $\Code$ is composed of cosets of the zero-error code $\Code^0=\{0,2,\ldots,q-2\}^n$. In particular, let
\begin{equation}
\Code=\Code^0+\Code_2
\label{eq:CfromC_2}
\end{equation}
where $\Code_2\subseteq \{0,1\}^n$ is a binary code and where the sum is the ordinary sum of vectors in $\Z_q^n$. It is easy to see that if $\Code_2$ is linear over $\Z_2$ then $\Code$ is linear over $\Z_q$. This is because $\Code^0$ is linear and the $q$-ary sum of two codewords in $\Code_2$ can be decomposed as the sum of a codeword in $\Code^0$ and a codeword in $\Code_2$. In this case then, for the spectrum components in \eqref{eq:Pefromspectrum} we have the simpler expression
\begin{equation*}
A_z=\left|\{i>1:  w(\bm{x}_i)=z\} \right|.
\end{equation*}
where $w(\bm{x}_i)=d(\bm{x}_i,\bm{0})$ is the weight of codeword $\bm{x}_i$ and we assume $\bm{x}_1=\bm{0}$.
We can now relate the spectrum of $\Code$ under the metric $d$ with the spectrum of $\Code_2$ under the usual Hamming metric.
We observe that any codeword of $\Code_2$ of Hamming weight $z$ leads to $2^z$ codewords of $\Code$ of weight $z$ and ${(q/2)}^n-2^z$ codewords of infinite weight. So, we can write 
\begin{equation}
A_z=2^z B_z\,,
\label{eq:fromAtoBspectrum}
\end{equation}
where $B_z$ is the number of codewords in $\Code_2$ of Hamming weight $z$. Let now $r$ be the rate of $\Code_2$. It is known from the Gilbert-Varshamov procedure (see for example \cite{poltyrev-1994} and \cite[Sec. II.C]{barg-forney-2002}) that, as $n\to\infty$,  binary linear codes of rate $r$ exist whose spectra satisfy
\begin{equation}
B_{\delta n}=
\begin{cases}
0 & \mbox{if } \delta < \delta_{GV}(r;2)\\
e^{n(r-\log(2)+h_2(\delta)+o(1))} & \mbox{if } \delta\geq  \delta_{GV}(r;2)
\end{cases}\,.
\label{eq:GVBinarySpectrum}
\end{equation}
Such binary codes of rate $r$ used in the role of $\Code_2$ in \eqref{eq:CfromC_2} lead to codes $\Code$ with rate $R=\log(q/2)+r$ whose error exponent can be bounded to the first order by the leading term in the summation \eqref{eq:Pefromspectrum}. Using  \eqref{eq:fromAtoBspectrum} and \eqref{eq:GVBinarySpectrum} we find
\begin{equation}
\frac{1}{n}\log \Pe  \leq \max_{\delta\geq \delta_{GV}(r;2)} \left[ r-\log(2) +h_2(\delta)+\delta \log(2 \aleps)\right]+o(1)\,.
\label{eq:Peasmaximum}
\end{equation}
The argument of the maximum is increasing for $\delta\leq 2\aleps/(1+2\aleps)$ where it achieves the maximum value
\begin{equation}
r-\log(2)+h_2\left(\frac{2\aleps}{1+2\aleps}\right)+\frac{2\aleps}{1+2\aleps}\log(2 \aleps)
\end{equation}
which, using $R=\log(q/2)+r$, simplifies to
\begin{equation}
R-\log\frac{q}{1+2\aleps}\,.\
\end{equation}
This is thus the maximum in \eqref{eq:Peasmaximum} if $2\aleps/(1+2\aleps)\geq \delta_{GV}(r;2)$ or, equivalently, if
\begin{align}
 R & \geq \log q - h_2\left(\frac{2 \aleps}{1+2\aleps}\right)\\
 & = \log\frac{q}{2} + R_{\text{ex}}^*(\epsilon;2).
\end{align}
Otherwise, the maximum in  \eqref{eq:Peasmaximum}  is achieved at $\delta=\delta_{GV}(r;2)$ and has value  $\delta_{GV}(r)\log 2 \aleps$.

So, we have the bound
\begin{equation}
E(R)\geq
\begin{cases}
\log\frac{2}{1+2\aleps}-(R-\log(\frac{q}{2})) ,&  R\geq \log\frac{q}{2} + R_{\text{ex}}^*(\epsilon;2)\\
-\delta_{GV}(R-\log\frac{q}{2};2)\log(2 \aleps) ,&  \log\frac{q}{2}< R\leq \log\frac{q}{2} + R_{\text{ex}}^*(\epsilon;2)\,.
\end{cases}
\end{equation}
The expression on the right hand side is simply $\Eex^{\text{BSC}}(R-\log(q/2))$ as defined in equation \eqref{eq:Eex_BSC}.
\end{IEEEproof}

\begin{figure}
\centering
\includegraphics[width=\linewidth]{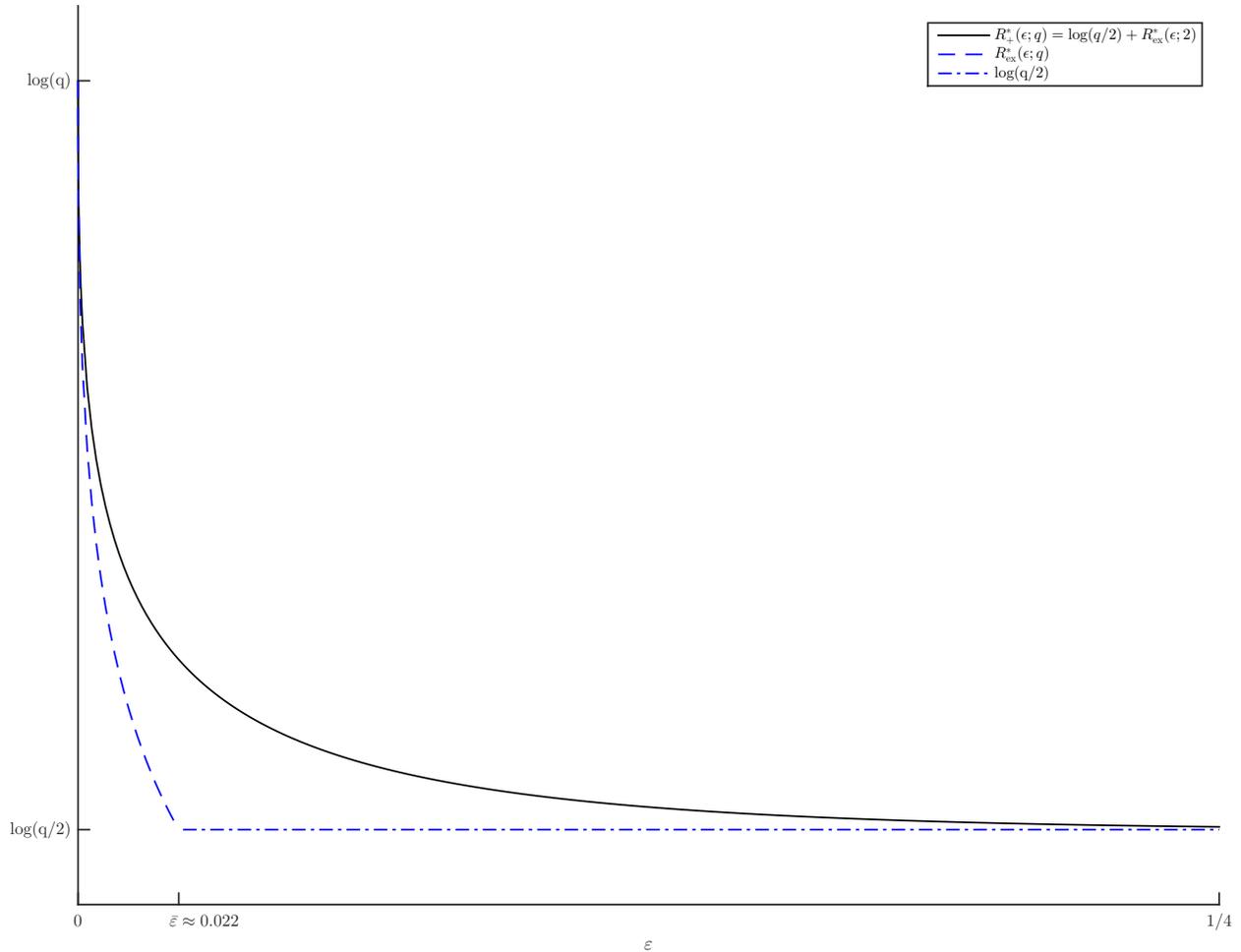}
\caption{A comparison of the rate values at which the expurgated bound and the new bound derived in Theorem \ref{th:GV-expurgatedbound} depart from the straight line of slope -1, for even values of $q\geq 4$. Note that for different such $q$'s the functions have the same shape and only shift vertically. In particular, from equation \eqref{eq:Eex_condbareps} we find that $\bar{\epsilon}\approx 0.022$ for all even $q\geq 4$.}
\label{fig:Rstar}
\end{figure}

\begin{figure}
\centering
\includegraphics[width=\linewidth]{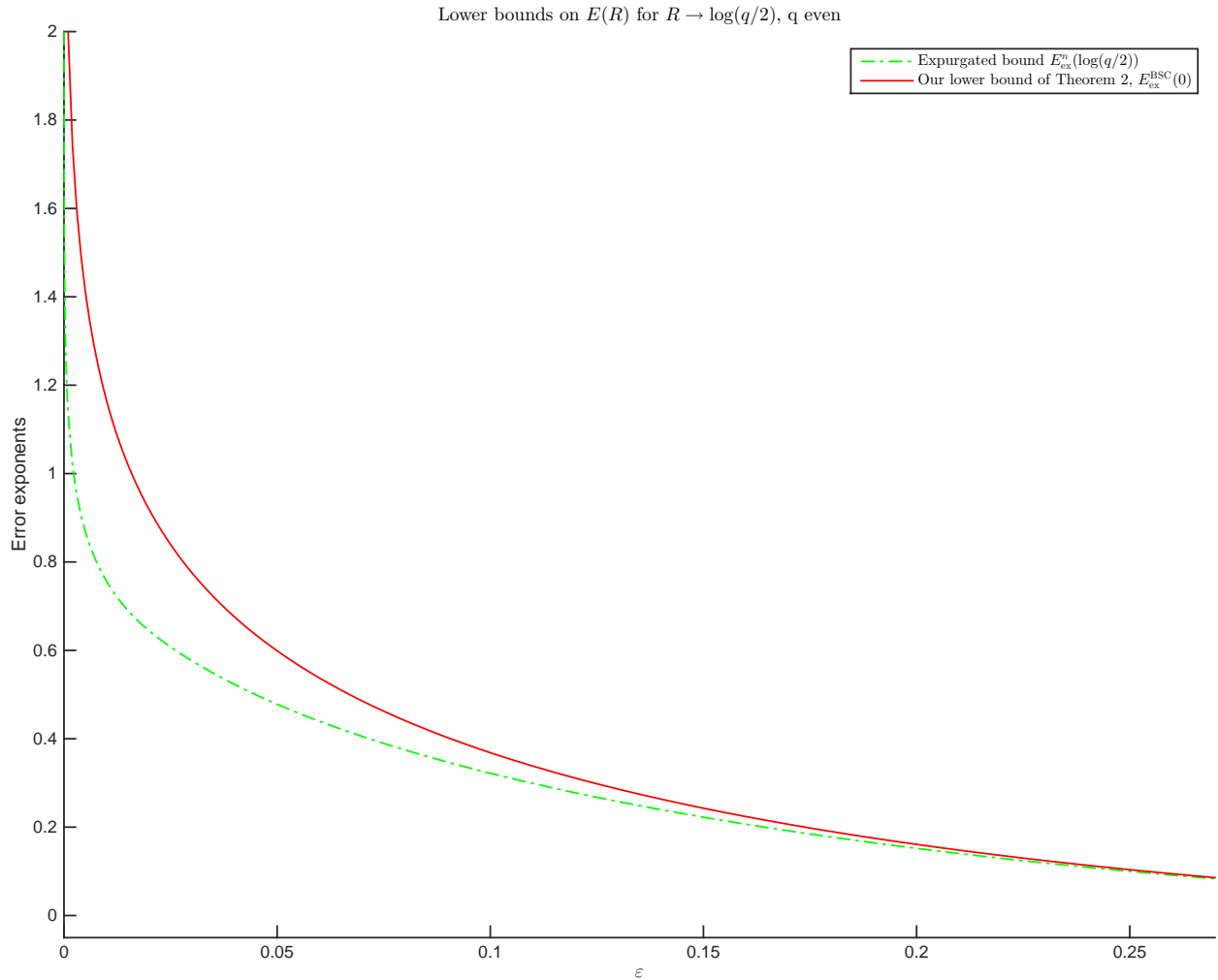}
\caption{A comparison of lower bounds on the value $E(R)$ for even $q$ as $R$ approaches the zero-error capacity $\log(q/2)$, as a function of the channel parameter $\epsilon$. Our lower bound of Theorem \ref{th:GV-expurgatedbound} is strictly better than the expurgated lower bound for all $\epsilon>0$. As $\epsilon\to 0$ the two bounds clearly diverge and their ratio tends to $(1+2\alpha_{\bar{\epsilon}})/(4\alpha_{\bar{\epsilon}})\approx 2.2017$.}
\label{fig:Emax}
\end{figure}

\begin{figure}
\centering
\includegraphics[width=0.9\linewidth]{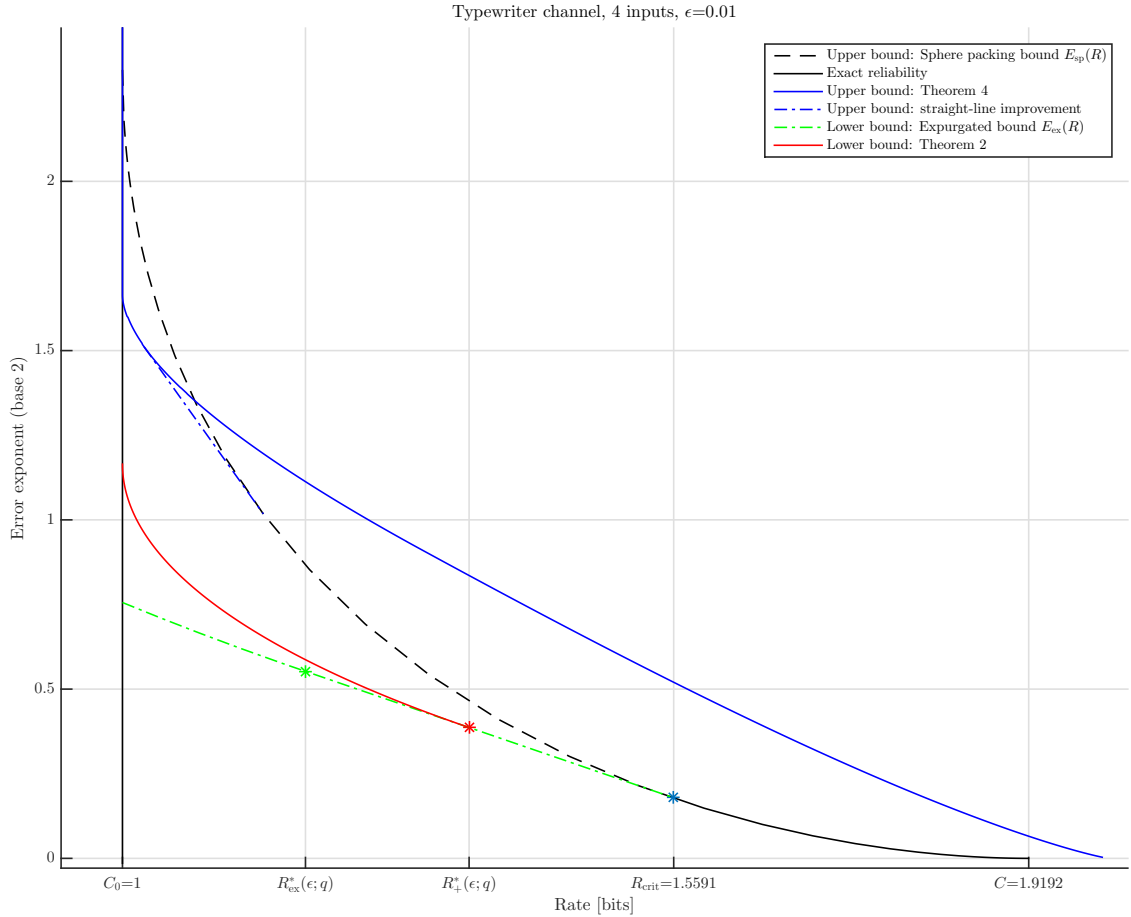}
\caption{Bounds on $E(R)$ for a typewriter channel with 4 inputs, $\epsilon=0.01$.}
\label{fig:even_GV}
\end{figure}

A graphical comparison of our bound and the standard expurgated bound is visible in Figure \ref{fig:even_GV} for $q=4$ and $\epsilon=0.01$. Note that the straight line portion of the bound coincides with a portion of the straight line in the standard expurgated bound as in \eqref{eq:Eexeven_str}. However, the rate value $R_{\text{ex}}^*(\epsilon;q)$ at which the standard expurgated bound $\Eex^n(R)$ departs from the straight line is strictly smaller than the value $R_+^*(\epsilon;q)=\log\frac{q}{2} + R_{\text{ex}}^*(\epsilon;2)$ ($q$ even) at which our bound does for all $0<\epsilon<1/2$. A comparison of these two quantities for different $\epsilon$ is given in Figure \ref{fig:Rstar}. Finally, Figure \ref{fig:Emax} shows a comparison of the lower bounds on $E(R)$ at rates near $\log(q/2)$, for even $q\geq 4$ and varying $\epsilon$, which shows that our bound is always strictly better than the standard expurgated bound.

\begin{remark}
It is a remarkable fact that our bound corresponds exactly to the expurgated bound of a binary symmetric channel with cross-over probability $\epsilon$ shifted by $\log(q/2)$ on the $R$ axis. On one hand, it is not very surprising that a bound for \emph{binary} codes shows up, given the construction we used in \eqref{eq:CfromC_2}. On the other hand, it is curious to observe that we obtain \emph{specifically the expression of the BSC} because the coefficient $2^z$ which relates $A_z$ to $B_z$ in \eqref{eq:fromAtoBspectrum} leads to the coefficient $2$ inside the logarithm in \eqref{eq:Peasmaximum}, thus replacing the quantity $\sqrt{\epsilon(1-\epsilon)}$ which has to be used for the typewriter channel with the quantity $2\sqrt{\epsilon(1-\epsilon)}$ which appears in the expurgated bound of the BSC. 
\end{remark}

\begin{remark}
We think it is reasonable to consider the bound given in Theorem \ref{th:GV-expurgatedbound} as the correct modification of the expurgated bound for these particular channels. It is interesting to observe that the derived bound does not really use constructions which are totally out of reach with the standard expurgated bound; the zero-error code used in \eqref{eq:CfromC_2} is in fact also ``found'' by the standard expurgated bound as shown in Section \ref{sec:class_bd}. However, this zero-error code shows up in the standard expurgated bound only at very low rates, specifically at $R<\log(q/2)$, while our procedure shows that it is useful even at higher rates.
It is rather natural to ask then how the expurgated bound should be modified in general to exploit, at a given rate $R$, zero-error codes which would usually appear in that bound only at lower rates. 
\end{remark}

\begin{remark}
\label{rem:gap_evenq}
In the same way as the bound in Theorem \ref{th:GV-expurgatedbound} is a $\log(q/2)$-shifted version of the expurgated bound for the
BSC, it was already observed after equations \eqref{eq:Er_typewriter_form} and \eqref{eq:esp_tw} that the random coding bound and the sphere packing bound are also
$\log(q/2)$-shifted versions of the ones for the BSC. In particular, we find that at rate
$\log(q/2)$ our lower bound has value precisely half the value of the sphere packing bound, as happens at $R=0$ for the
BSC. However, while closing the gap at $R=0$ for the BSC is essentially trivial, for the typewriter channel it seems to
be a harder problem. See Section \ref{sec:class_bd} and Remark \ref{rem:spectrum_evenq} in Section \ref{sec:new_UP_spectrum}.
\end{remark}

For odd values of $q$, deriving a corresponding lower bound on $E(R)$ is difficult in general, since general good zero-error codes are not known or, in any case, have a rather complicated structure. One particular exception is the case $q=5$, for which an asymptotically optimal zero-error code is known.

\begin{theorem}
\label{th:GV-lower-5}
For $q=5$, in the range 
\begin{equation}
\log\sqrt{5}\leq R \leq \log R_+^*(\epsilon;5)
\end{equation}
where
\begin{equation}
R_+^*(\epsilon;5)=5 -\frac{1}{2}h_5\left(1-\frac{1}{(1+2\aleps)^2}\right)
\end{equation}
we have the lower bound
\begin{equation*}
E(R)\geq 
-\frac{1}{2}\delta_{GV}(2R -\log5;5)\log(\aleps(1+\aleps))\,.
\end{equation*}
\end{theorem}

\begin{figure}
\centering
\includegraphics[width=\linewidth]{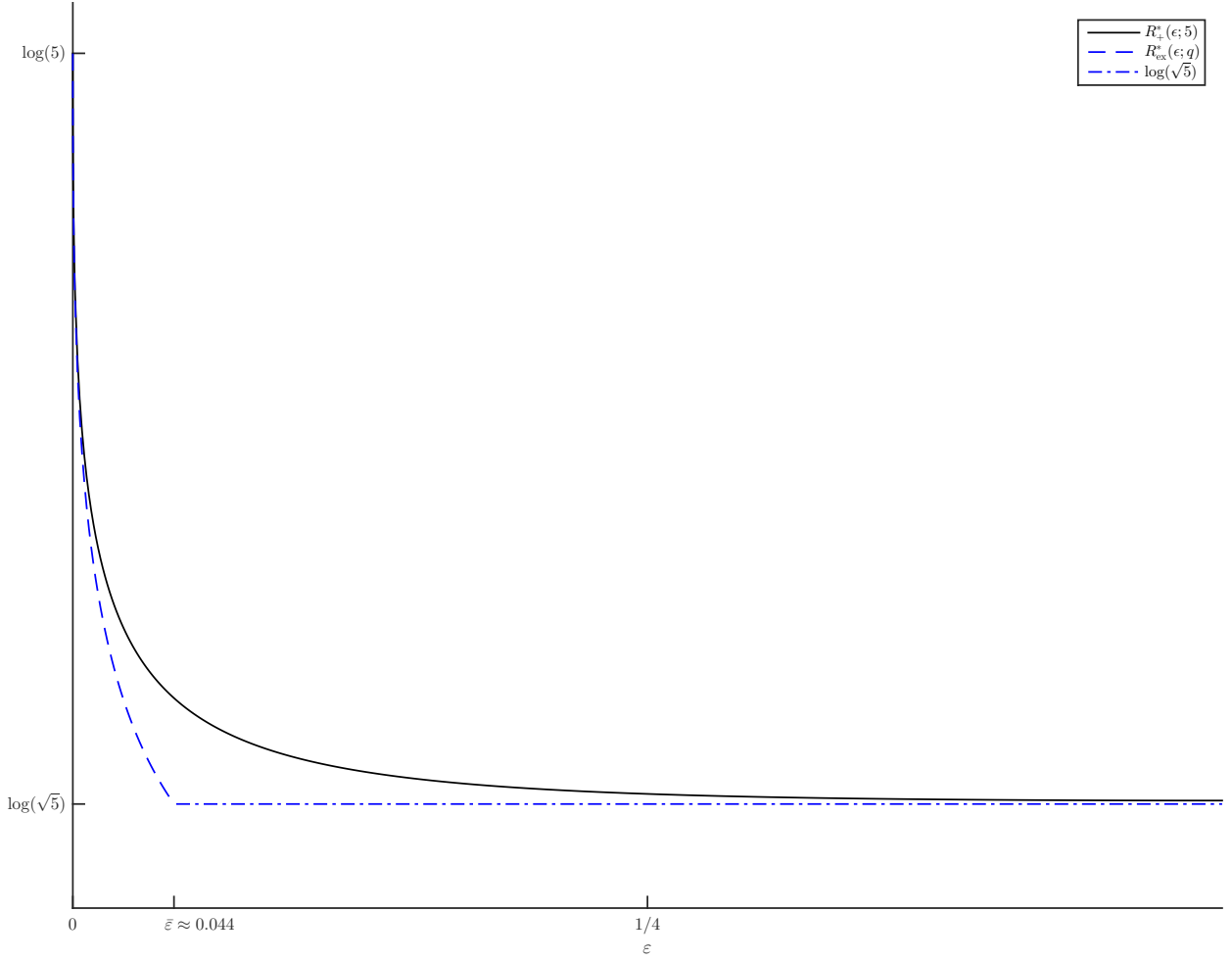}
\caption{A comparison of the rate values at which the expurgated bound and the new bound derived in Theorem \ref{th:GV-lower-5} depart from the straight line of slope -1 for $q=5$.}
\label{fig:Rstar5}
\end{figure}

\begin{figure}
\centering
\includegraphics[width=\linewidth]{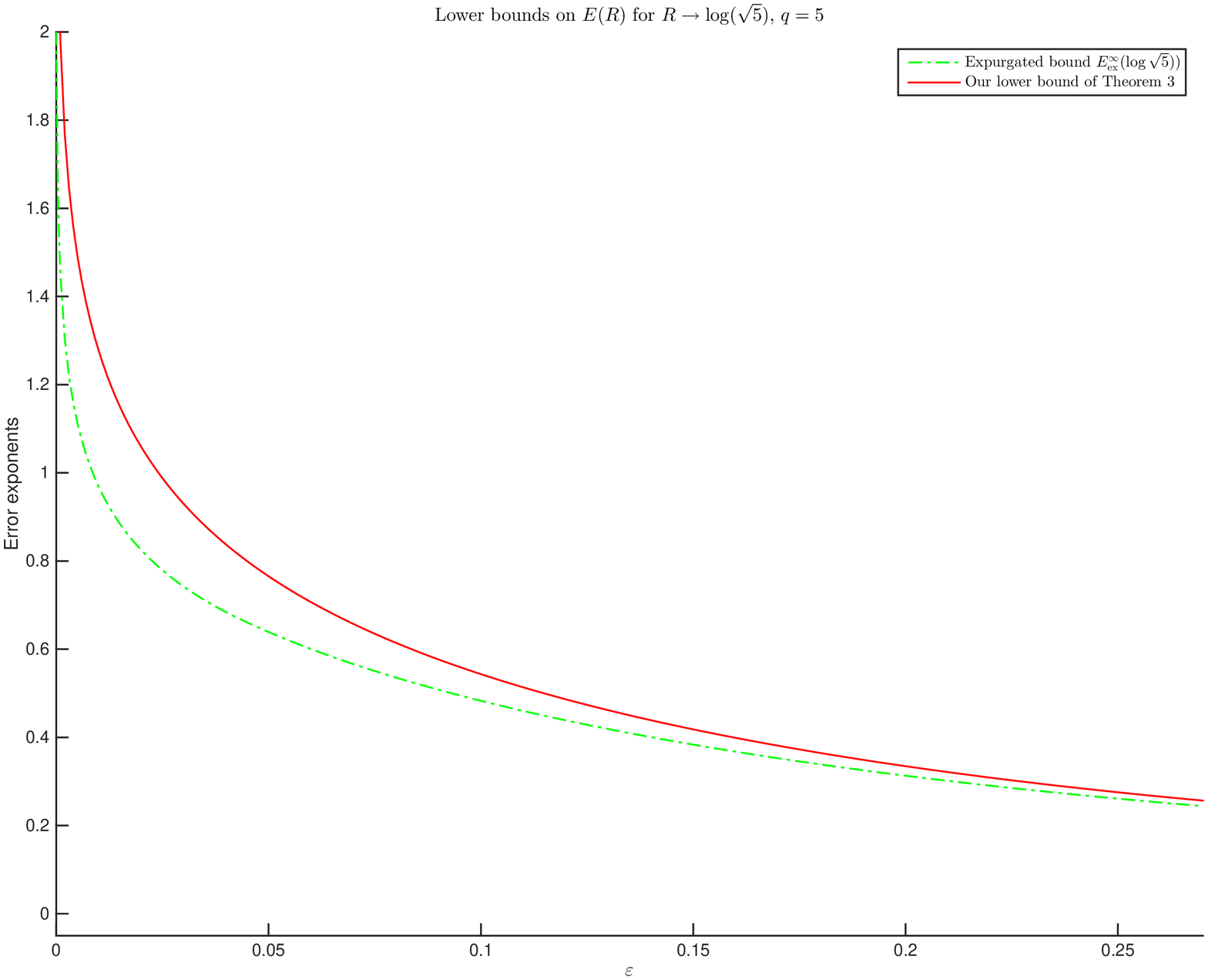}
\caption{A comparison of lower bounds on the value $E(R)$ for $q=5$ as $R$ approaches the zero-error capacity $\log(\sqrt{5})$, as a function of the channel parameter $\epsilon$. Our lower bound of Theorem \ref{th:GV-lower-5} is strictly better than the expurgated lower bound for all $\epsilon>0$. As $\epsilon\to 0$ the two bounds clearly diverge and their ratio tends to $(1+2\alpha_{\bar{\epsilon}})/(5\alpha_{\bar{\epsilon}})\approx 1.3752$.}
\label{fig:Emax5}
\end{figure}

\begin{figure}
\centering
\includegraphics[width=0.9\linewidth]{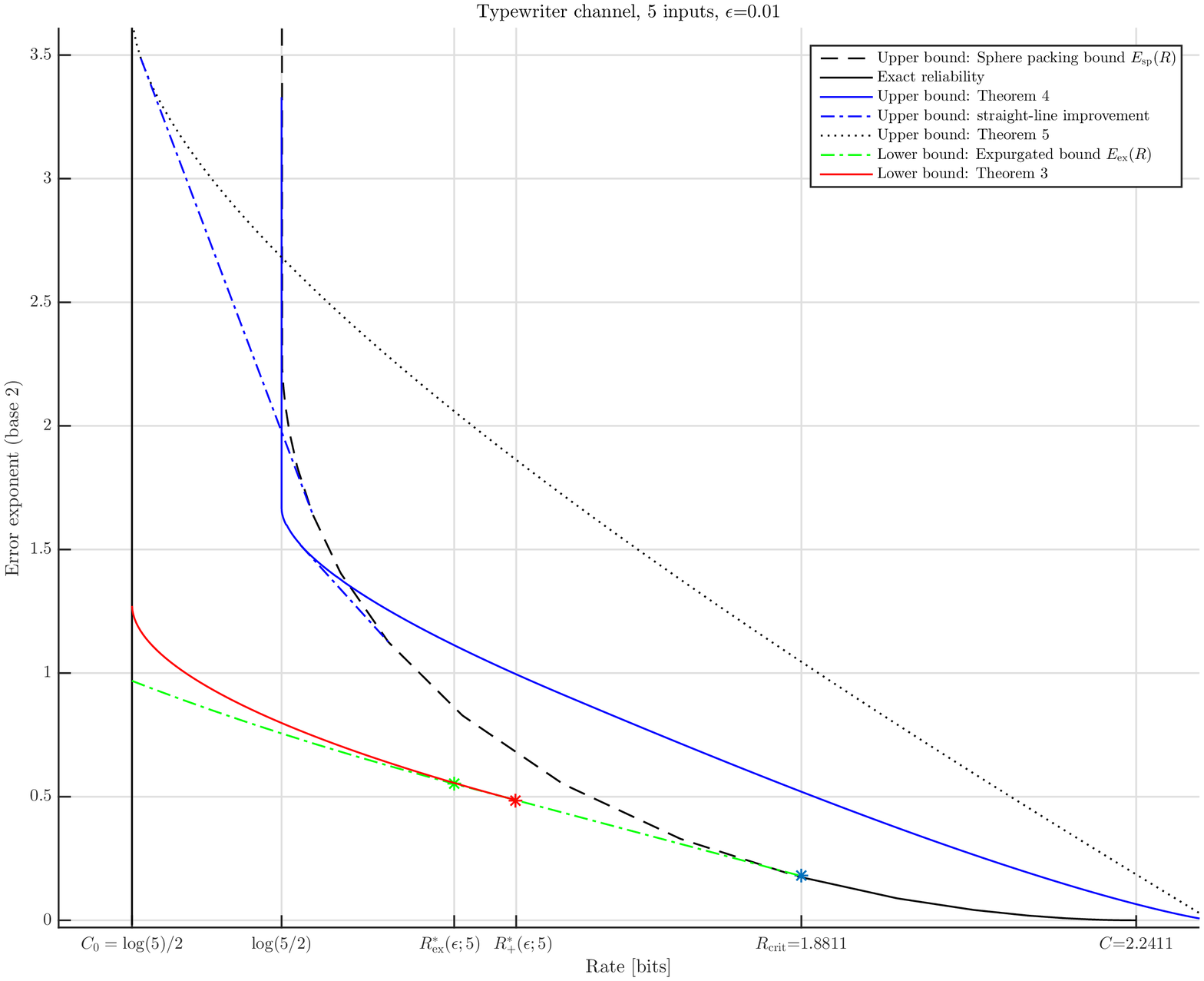}
\caption{Bounds on $E(R)$ for a typewriter channel with 5 inputs, $\epsilon=0.01$.}
\label{fig:q5_GV}
\end{figure}

A comparison of this bound with the expurgated bound is shown in Figure \ref{fig:q5_GV}. Note that at the upper extreme of the interval considered in Theorem {th:GV-lower-5}, $R=R_+^*(\epsilon;5)$, the given bound touches the expurgated bound of Theorem  \ref{th:exinf}. A comparison of this quantity with $R_{\text{ex}}^*(\epsilon;5)$ is shown in Figure \ref{fig:Rstar5}. Figure \ref{fig:Emax5} shows a comparison of the new bound with the expurgated bound as $R$ approaches the zero-error capacity $\log\sqrt{5}$.

\begin{IEEEproof}[Proof of Theorem \ref{th:GV-lower-5}]
We start from equation \eqref{eq:Pefromspectrum}, but restated for codes of even length $n'=2n$. In particular, consider linear codes with a $(n+k)\times 2n$ generator matrix of the form
\begin{equation*}
G^+=\left( 
\begin{array}{cc}
I_{n} &  2I_{n}\\
0 & G
\end{array}\right)
\end{equation*}
where $I_n$ is the $n\times n$ identity matrix and $G$ is a $k\times n$ matrix of rank $k$ over $\Z_5$. Note that this corresponds to taking $5^k$ cosets of the $n$-fold cartesian power of Shannon's zero-error code of length 2 \cite{shannon-1956}.
Since we focus again on linear codes, the $A_z$'s in \eqref{eq:Pefromspectrum} still take the simple form $A_z = \left|\{i>1 :  w(\bm{x}_i)=z\} \right|$.

We now proceed to the study of $A_z$.
We can decompose any information sequence $\bm{u}\in \mathbb{Z}_5^{n+k}$ in two parts, $\bm{u}=(\bm{u}_1, \bm{u}_2)$, with $\bm{u}_1 \in \mathbb{Z}_5^n$ and $\bm{u}_2\in\mathbb{Z}_5^k$. The associated codeword $\bm{v}=\bm{u}G^+$ can be correspondingly decomposed in two parts $\bm{v}=(\bm{v}_1, \bm{v}_2)$ with $\bm{v}_1=\bm{u}_1$ and $\bm{v}_2=2\bm{u}_1+\bm{u}_2G$. Call $\bm{\nu}=\bm{u}_2G$. We now relate the weight $w(\bm{v})$ to the \emph{Hamming weight} $\wH(\bm{\nu})$ and to the form of $\bm{u}_1$.
Note in particular that we can write
\begin{equation*}
w(\bm{v})=\sum_{i=1}^n w((v_{1,i},v_{2,i}))
\end{equation*}
and that
\begin{equation*}
(v_{1,i},v_{2,i})=u_{1,i}(1,2)+(0,\nu_i)\,.
\end{equation*}

Note first that $w(\bm{v})=\infty$ if $u_{1,i}=\pm 2$ for some $i$. So, for the study of $A_z$ we need only consider the cases $u_{1,i}\in \{0,\pm 1\}$. Consider first the case when $\nu_{i}=0$. If $u_{1,i}=0$ then $w(v_{1,i})=w(v_{2,i})=0$ while if $u_{1,i}=\pm 1$ then $w(v_{2,i})$ is infinite. So, if $\nu_{i}=0$ one choice of $u_{1,i}$ gives no contribution to $w(\bm{v})$ while all other choices lead to $w(\bm{v})=\infty$, and hence give no contribution to $A_z$ for any finite $z$.
Consider then the case of a component $\nu_{i}\neq 0$. It is not too difficult to check that one choice of $u_{1,i}$ in $\{0,\pm 1\}$ gives $w(v_{1,i})=w(v_{2,i})=1$, one gives $w(v_{1,i})=1$ and $w(v_{2,i})=0$ or vice-versa, and the remaining one gives  $w(v_{2,i})=\infty$.
So, if $\nu_{i}\neq 0$ one choice of $u_{1,i}$ contributes $1$ to $w(\bm{v})$, one choice of $u_{1,i}$ contributes $2$, while all other choices lead to $w(\bm{v})=\infty$, and hence give no contribution to $A_z$ for any finite $z$.

So, for a fixed $\bm{\nu}$ of Hamming weight $d$, and for a fixed $t\in\{1,2,\ldots,d\}$, there are $\binom{d}{t}$ vectors $\bm{u}_1$ which give codewords $\bm{v}$ of weight $2t+(d-t)=d+t$. If $B_d$ is the number of sequences $\bm{u}_2$ which lead to a $\bm{\nu}$ of Hamming weight $d$, then we have
\begin{equation}
\Pe\leq \sum_{d=1}^n \sum_{t=1}^d B_d \binom{d}{t} \aleps^{(d+t)}\,.\label{eq:SumErLB}
\end{equation}
But $B_d$ is now simply the Hamming spectrum component of the linear code with generator matrix $G$, and it is known (see \cite[Prop. 1]{katsman-tsfasman-vladut-1992}) that as we let $n$ and $k$ grow to infinity with ratio $(k/n) \log 5\to r$, matrices $G$ exist for which
\begin{equation*}
B_{\delta n}=
\begin{cases}
0 & \mbox{if } \delta < \delta_{GV}(r;5)\\
e^{n(r-\log 5+h_2(\delta) + \delta \log(4)+o(1))} & \mbox{if } \delta\geq  \delta_{GV}(r;5)
\end{cases}\,.
\end{equation*}
Defining $\delta=d/n$, $\tau=t/d$ and $r=(k/n)\log(5)$, the probability of error is bounded to the first order in the exponent by the largest term in the sum \eqref{eq:SumErLB} as
\begin{equation*}
\frac{1}{2n}\log \Pe  \leq \frac{1}{2}\cdot \max_{\delta\geq \delta_{GV}(r;5), \tau\in[0,1]} [r-\log(5)+h_2(\delta)+2\delta\log 2 +\delta h_2(\tau)+(\delta+\delta\cdot\tau)\log\aleps] + o(1)\,.
\end{equation*}
The maximum over $\tau$ is obtained by maximizing $h_2(\tau)+\tau \log \aleps$, which is solved by $\tau=\aleps/(1+\aleps)$ with maximum value $\log(1+\aleps)$, independently of $\delta$. So, we are left with the maximization
\begin{equation*}
\max_{\delta\geq \delta_{GV}(r;5)} \left[r-\log(5)+h_2(\delta)+\delta\log \beta \right]\,,\qquad \beta=4\aleps(1+\aleps)\,.
\end{equation*}
The argument is increasing for $\delta\leq \beta/(1+\beta)$, where it achieves the maximum value  $r-\log(5)+\log(1+\beta)$, and decreasing for larger values of $\delta$. So, the maximizing $\delta$ is $\delta=\beta/(1+\beta)$ if $\beta/(1+\beta)\geq\delta_{GV}(r;5)$ and $\delta_{GV}(r;5)$ otherwise.
Combining these facts, noticing that $1+\beta=(1+2\aleps)^2$, we find
\begin{equation*}
\frac{1}{2n}\log\Pe\leq \frac{1}{2}
\begin{cases}
(r-\log(5)+2\log(1+2\aleps)) +o(1),& \delta_{GV}(r;5)\leq \beta/(1+\beta)\\
\delta_{GV}(r;5)\log(\aleps(1+\aleps)) +o(1),&  \delta_{GV}(r;5)> \beta/(1+\beta) \,.
\end{cases}
\end{equation*}
Considering that the block length is $2n$ and the rate of the global code is $R=(\log(5)+r)/2$ with $r\geq 0$, after some simple algebraic manipulations we obtain 
\begin{equation*}
E(R)\geq
\begin{cases}
\log \frac{5}{1+2\aleps}-R,& R\geq \log 5 -\frac{1}{2}h_5\left(1-\frac{1}{(1+2\aleps)^2}\right)\\
-\frac{1}{2}\delta_{GV}(2R -\log5;5)\log(\aleps(1+\aleps)),& \log\sqrt{5}\leq R< \log 5 -\frac{1}{2}h_5\left(1-\frac{1}{(1+2\aleps)^2}\right)\,.
\end{cases}
\end{equation*}
The first part of the bound coincides with the standard straight line portion of the expurgated bound, while the second part is the claimed new bound.
\end{IEEEproof}

\section{New upper (converse) bounds on $E(R)$}
\label{sec:UPs}

We present three different new upper bounds below, each of which is the tightest known bound for certain values of
$q,R,\epsilon$.

\subsection{Bound via a reduction to binary codes}\label{sec:UPs_binary}

We have already evaluated the sphere-packing bound above~\eqref{eq:esp_tw}. For channels with $C_0=0$, the sphere packing bound is known to be weak at low rates. In particular, it was proved by Berlekamp \cite{berlekamp-thesis} \cite{shannon-gallager-berlekamp-1967-2} that the expurgated bound is tight at $R=0+$ for all channels with $C_0=0$. For general channels with positive $C_0$ no similar result is known. In the case of typewriter channels, standard methods can be adapted to give the following result.

\begin{theorem}
For $R>\log(q/2)$ and any $q \ge 4$ we have
\begin{equation}
E(R)\leq \delta_{LP2}(R - \log(q/2)) \log {1\over \aleps}\,,
\end{equation}
where $\delta_{LP2}(\cdot)$ is the second MRRW linear-programming bound \cite{mceliece-et-al-1977} for the binary Hamming space given by
$$ \delta_{LP2}(R) = \min 2{\alpha (1-\alpha) - \beta(1-\beta)\over 1+2\sqrt{\beta (1-\beta)}}\,,
$$
where the minimum is over all $0 \le \beta \le \alpha \le {1\over 2}$ satisfying $\log 2- h_2(\alpha) + h_2(\beta) \ge R$.
\label{Th:binary_JPL2}
\end{theorem}
\begin{IEEEproof}
The key point is to observe that any code $\Code$ of rate $R>\log(q/2)$ admits a subcode of rate at least $R-\log(q/2)$
whose codewords are all pairwise confusable. This is can be proved in the following way\footnote{In graph theoretic
terms, this is proved by observing that the fractional clique covering number of the cycle of length $q$ is $q/2$. }.
For a sequence $v\in \Z_q^n$ let $\Code_v$ be the set of codewords whose $i$-th component is in the set
$\{v_i,v_i+1\}\mod n$, for all $i\in\{1,2\ldots,n\}$. Observe that all the codewords in $\Code_v$ are pairwise confusable. Let then
$V=(V_1,V_2,\ldots,V_n)$ be an i.i.d. sequence of uniform random variables in $\Z_q$. The expected size of $\Code_V$ is
$|\Code|(2/q)^n$ and if $|\Code|=e^{nR}$ with $R>\log(q/2)$ then for at least one $v$ we have $|\Code_v|\geq
e^{n(R-\log(q/2))}$.
Fix this particular $v$ and consider only the sub-code $\Code_v$, whose probability of error cannot be larger than that
of $\Code$. Since in any coordinate all codewords $\Code_v$ use only the same two (confusable) symbols, using $\Code_v$ on the typewriter channel is equivalent to using a binary code on a binary
\emph{asymmetric} erasure channel. Then from $\Code_v$ we may select a full-rate ``constant-composition'' subcode $\Code_v'$, in the sense that its binary equivalent representation is a constant composition binary code. It is well known,
e.g.~\cite[Ex. 10.20c]{csiszar-korner-bookv2}, that for such a subcode $\Code_v'$ the probability of error is lower-bounded
by
$$ P_e(\Code_v') \ge \alpha_\epsilon^{d_{min}(\Code_v') + o(n)}\,,$$
where $d_{\min}(\Code_v')$ is the minimal Hamming distance of $\Code_v'$. (Note that here we also used the explicit
expression for the Bhattacharya distance as in~\eqref{eq:bhatta}.) From~\cite{mceliece-et-al-1977} it follows
that 
$$ d_{min}(\Code_v') \le n \delta_{LP2}(R-\log(q/2)) + o(n)\,.$$
\end{IEEEproof}

We observe that Theorem \ref{Th:binary_JPL2} implies 
\begin{equation}
E(\log(q/2)+\delta)\leq \frac{1}{2}\log\frac{1}{\aleps}\,, \qquad \forall \delta>0
\label{eq:logq/2-bound}
\end{equation}
which improves the sphere packing bound $\Esp(\log(q/2))=-\log(1/2\aleps)$ for $\epsilon< 1/2-\sqrt{3}/4\approx~0.067$. In this case one can combine Theorem \ref{Th:binary_JPL2} with the straight line bound, which asserts that any line connecting a point at $R_1$ on the curve of a low-rate upper bound on $E(R)$ to a point on the curve of the sphere packing bound at a higher rate $R_2$ is also an upper bound on $E(R)$ in the intermediate range of rate values $R_1\leq R \leq R_2$.

It is worth discussing \eqref{eq:logq/2-bound} in more detail for even values of $q$. In this case, in fact, the lower bound we derived in Section \ref{sec:new_LB} is a $\log(q/2)$-shifted version of the expurgated bound for the BSC, and the same is true for the sphere packing bound. Given the structure of the proof, one might then expect that our upper bound of Theorem \ref{Th:binary_JPL2} should match the lower bound of Theorem \ref{th:GV-expurgatedbound} as happens at $R=0$ for the BSC.
The comment in the previous paragraph implies that this is not the case; it is actually easy to check that the bounds are off by $\log \sqrt{2}$ at the rates near $\log(q/2)$, and the reader might wonder what is happening. Implicit in the proofs of Theorems \ref{th:GV-expurgatedbound} and \ref{Th:binary_JPL2} (see also \cite[Prop. 1]{dalai-polyanskiy-2015a} or \cite[Prop. 5]{cullina-dalai-polyanskiy-2016}) is the fact that, for even $q$, the largest possible minimum distance (with our semimetric) of codes at rates near $\log(q/2)$ is known and it is precisely $1/2$. So one might expect that the exact value of $E(\log(q/2)^+)$ should also be known. The main reason why this is not so is the possible presence of many codeword pairs at minimum distance.
Our procedure in Theorem \ref{Th:binary_JPL2} gives a \emph{upper} bound on $E(R)$ by lower bounding the probability of error in a binary hypothesis test between \emph{two} codewords at minimum distance. However, our \emph{lower} bound on $E(R)$ in Theorem \ref{th:GV-expurgatedbound} uses codes such that any codeword has $2^d$ neighboring codewords at minimum distance $d$. When plugged in the union bound, this leads to the mentioned gap between lower and upper bounds on $E(R)$.
Note in particular that this effect is related to the coefficient $2^z$ which relates $A_z$ to $B_z$ in \eqref{eq:fromAtoBspectrum} already mentioned at the end of the proof of Theorem \ref{th:GV-expurgatedbound}.
Since we have no insight on possible improvements of Theorem \ref{th:GV-expurgatedbound} while we do know of cases where Theorem \ref{Th:binary_JPL2} is even weaker than the sphere packing bound, we deduce that Theorem \ref{Th:binary_JPL2} is weak because it fails to catch the possible presence of a high number of neighbors. We will see later (see Remark \ref{rem:spectrum_evenq} in Section \ref{sec:new_UP_spectrum}) that a different procedure does allow one to prove the presence of an exponential number of neighbors, though it turns out to be difficult to convert this into a good bound on the probability of error for the asymmetric case of $\epsilon\neq 1/2$, which is unfortunately the only interesting one for even values of $q$.

We finally comment on the optimal use of the straight line bound for odd $q$ using known results in the literature.
The sphere packing bound and  Theorem \ref{Th:binary_JPL2} are only useful at rates larger than $\log(q/2)$. However, for odd $q$, Lov\'asz's bound on the zero-error capacity implies that $E(R)$ is finite for all rates $R>\log \theta(C_q) $,
where $\theta(C_q) $ is the Lov\'asz theta function of equation  \eqref{eq:lovasz_theta_cq}.
It is then possible to use the straight line bound in a rather simple way. Any code of rate $R>\log \theta(C_q)$ has a positive \emph{maximal} probability of error, and this cannot be smaller than $\epsilon^n$, since $\epsilon$ is the smallest non-zero transition probability. Hence, the reliability function satisfies $E(R)\leq -\log \epsilon$ for $R>\log \theta(C_q)$ and application of the straight line bound gives the following result.
\begin{theorem}
For odd $q$, the segment connecting the point $(\log \theta(C_q), \log 1/\epsilon)$ tangentially to the sphere packing curve in the $(R,E)$ plane is an upper bound on $E(R)$.
\end{theorem}

\subsection{Bound via minimum distance}
\label{sec:new_UP_distance}

In this Section we present a new upper bound on $E(R)$ for typewriter channels based on the minimum distance of codes. Our focus here is on the case where $q$ is odd. Furthermore, although we state the bound for general $\epsilon$, it is developed with a main focus on the symmetric case $\epsilon=1/2$. 
In the next section we will derive an improved version of the bound which only holds for $\epsilon=1/2$. We believe the case $\epsilon< 1/2$ would be worth attention for the extension of that bound rather than the optimization of the current one.

The main contribution of the bound presented here relies on combining bounds on the zero-error capacity with bounds on minimum distance of codes, which allows us to derive a new upper bound on $E(R)$ for rates larger than Lov\'asz's upper bound on $C_0$. More specifically, our bound derives from  an instance of the Delsarte linear programming  bound \cite{delsarte-1973} which combines the construction used by Lov\'asz for bounding the graph capacity \cite{lovasz-1979} with the construction used in \cite{mceliece-et-al-1977} to bound the minimum distance of codes in Hamming spaces (see \cite{schrijver-1979} and \cite{mceliece-et-al-1978} for discussions on the connection between Delsarte's and Lov\'asz's bounds).

\begin{theorem}
\label{th:ELP}
For odd $q$ and any $\delta \in (0,1)$ we have $E(R) \le \delta \log{1\over \epsilon}$ whenever
	\begin{align}	
		R &\ge \log\theta(C_q)+R_{\text{LP1}}\left(\frac{q}{\theta(C_q)}, \delta\right)\,,
	\end{align}	
where
\begin{align*}
R_{\text{LP1}}(q',\delta) &= h_{q'}\left({(q'-1)-(q'-2)\delta - 2\sqrt{(q'-1)\delta(1-\delta)}\over q'}\right)
\end{align*}
is the linear programming bound for codes in a $q'$-ary Hamming space \cite{mceliece-et-al-1977,aaltonen1990new} (with 
$q'$ not necessarily integral) and $\theta(C_q)=  q/(1+\cos(\pi/q)^{-1})$ is the Lov\'asz $\theta$-function for the
$q$-cycle $C_q$.
\end{theorem}

\begin{IEEEproof}
We lower bound the maximal probability of error over all codewords $\Pemax$, which in turn we bound in terms of an upper bound on the minimum distance of codes for the distance measure introduced in the previous section. Note in particular that we have 
\begin{equation*}
\Pemax\geq \max_{i\neq j}\frac{1}{2} \cdot \epsilon^{d(\bm{x}_i,\bm{x}_j)}\,.
\end{equation*}
Indeed, if there is no pair of confusable codewords, then the inequality is trivial. If instead codewords $i$ and $j$ are confusable, then they share a common output sequence which can be reached with probability at least $\epsilon^{d(\bm{x}_i,\bm{x}_j)}$ by both input $i$ and $j$; upon receiving it, any (possibly randomized) decoder will decode in error with probability at least $1/2$ either when codeword $i$ or codeword $j$ is sent. 
So, we can bound the reliability as
\begin{equation}
E(R)\leq \min_{i\neq j} \frac{1}{n}d(\bm{x}_i,\bm{x}_j) (1+o(1))\log(1/\epsilon)\,.
\label{eq:Etod}
\end{equation}
The rest of this section is devoted to bounding the minimum distance. In particular we prove that codes for which 
\begin{equation*}
 \min_{i\neq j} \frac{1}{n}d(\bm{x}_i,\bm{x}_j) \geq \delta
\end{equation*}
have rate $R$ upper bounded as
\begin{equation}
R\leq  \log \theta(C_q) + R_{\text{LP1}}\left(\frac{q}{\theta(C_q)}, \delta\right)(1+o(1))\,.
\label{eq:R_LP'_RLP}
\end{equation}
Note that Theorem \ref{th:ELP} follows from equations \eqref{eq:Etod}-\eqref{eq:R_LP'_RLP}. 

Our bound is based on $\theta$ functions and Delsarte's linear programming bound \cite{delsarte-1973}, but it is easier to describe it in terms of Fourier transforms. 
For any 
$f:\mathbb{Z}_q^n\to \mathbb{C}$ we define its Fourier transform as
\[
\hat f(\bm{\omega}) = \sum_{\bm{x}\in \mathbb{Z}_q^n} f(\bm{x})e^{\frac{2\pi i}{q}<\bm{\omega},\bm{x}>},\quad \bm{\omega}\in \mathbb{Z}_q^n
\]
where the non-degenerate $\mathbb{Z}_q$-valued bilinear form is defined as usual
$$<\bm{x},\bm{y}>\eqdef \sum_{k=1}^n x_k y_k\,. $$ 
We also define the inner product as follows
$$ (f,g) \eqdef q^{-n} \sum_{\bm{x}\in \mathbb{Z}_q^n} \bar f(\bm{x}) g(\bm{x})\,. $$

The starting point is a known rephrasing of linear programming bound. Let $\matc$ be a code with minimum distance at
least $d$. Let $f$ be such that $f(\bm{x})\leq 0$ if $d(\bm{x},\bm{0})=w(\bm{x})\geq d$, $\hat{f}\geq 0$ and $\hat{f}(\bm{0})>0$. Then,  consider the Plancherel  identity
\begin{equation}
q^n(f* 1_\matc, 1_\matc)=(\hat{f}\cdot\widehat{1_\matc}, \widehat{1_\matc})\,,
\label{eq:Plancharel}
\end{equation}
where $1_A$ is the indicator function of a set $A$. Upper bounding the left hand side by $|\matc|f(\bm{0})$ and lower
bounding the right hand side by the zero-frequency term $q^{-n}\hat{f}(\bm{0})|\matc|^2$, one gets
\begin{align}\label{eq:f0fhat0}
|\matc| & \leq \min q^n \frac{f(\bm{0})}{\hat{f}(\bm{0})}.
\end{align}
The proof of our theorem is based on a choice of $f$ which combines Lov\'asz' assignment used to obtain his bound on the zero-error capacity with the one used in \cite{mceliece-et-al-1977} to obtain bounds on the minimum distance of codes in Hamming spaces.

Observe first that Lov\'asz assignment can be written in one dimension ($n=1$) as
\[
g_1(x) = 1_0(x)+ \varphi  1_{\pm 1} (x),\quad x\in \mathbb{Z}_q\,,
\]
where  $\varphi=(2\cos(\pi/q))^{-1}$. Note that this function $g_1$ actually corresponds to the first row of the matrix $g_1^{\odot 1/\rho}$ defined in \eqref{eq:g1_1/rho} computed for $\rho=\bar{\rho}$, the largest value of $\rho$ for which the matrix is positive semi-definite. In the Fourier domain we thus have
\begin{equation*}
\widehat{g_1}(\omega)=1+2\varphi \cos(2 \pi \omega/q), \quad \omega\in \mathbb{Z}_q\,,
\end{equation*}
which satisfies $\widehat{g_1}\geq 0$ and, additionally, $\widehat{g_1}(\omega)=0$ for $\omega=\pm c$, with $c=(q-1)/2$. 
Correspondingly, define the $n$-dimensional assignment 
\[
g(\bm{x}) = \prod_{j=1}^n g_1(x_j), \quad 
\hat g(\bm{\omega}) = \prod_{j=1}^n\widehat{g_1}(\omega_j), \quad \bm{x},\bm{\omega} \in \mathbb{Z}_q^n.
\]
So, $\hat{g}\geq 0$, with $\hat{g}(\bm{\omega})=0$ if $\bm{\omega}$ contains any $\pm c$ entry.
Since $g(\bm{x})=0$ for $\bm{x}\notin \{0,\pm 1\}^n$, $g$ satisfies all the properties required for $f$ in the case  $d=\infty$, and when used in place of $f$ in \eqref{eq:f0fhat0} it gives Lov\'asz' bound
\begin{align*}
|\matc| & \leq q^n\frac{g(\bm{0})}{\hat{g}(\bm{0})}\\
& = (\theta(C_q))^n\\
& = q^n \left(\frac{\cos(\pi/q)}{1+\cos(\pi/q)}\right)^n
\end{align*}
for codes of infinite minimum distance. Note that we also indirectly obtained this conclusion when bounding the
expurgated bound for odd values of $q$ in Theorem~\ref{th:exinf}, see Remark \ref{rem:on_expurgated_lovasz}.

For the case of finite $d\leq n$, we build a function $f$ of the form $f(\bm{x})=g(\bm{x})h(\bm{x})$, for an appropriate $h(\bm{x})$. In
particular, since $g(\bm{x})$ is non-negative and already takes care of setting $f(\bm{x})$ to zero if $\bm{x}\notin \{0,\pm1\}^n$, it suffices to choose $h$ such that $h(\bm{x})\leq 0 $ whenever $\bm{x}\in\{0,\pm1\}^n$ contains at least $d$ entries with value $\pm1$. We restrict attention to $h$ such that $\hat{h}\geq 0$, so that $\hat{f}=q^{-n}\hat{g}*\hat{h}\geq 0$. In particular, we consider functions $h$ whose Fourier transform is constant on each of the following ``spheres'' in $\mathbb{Z}_q^n$ 
\[
S_\ell^c = \{\bm{\omega}:|\{i:\omega_i=\pm c \}|=\ell,\;  |\{i:\omega_i=0 \}|=n-\ell\}\,,\quad \ell=0,\ldots, n\,,
\]
and zero outside. This choice is motivated by the fact, observed before, that $\hat g_1(\pm c)=0$. Restricting $\hat{h}$ to be null out of these spheres simplifies the problem considerably.
We thus define
\begin{equation}\label{eq:formofh}
\hat{h}(\bm{\omega})=\sum_{\ell=0}^n \hat{h}_\ell 1_{S_{\ell}^c}(\bm{\omega})\,,\quad h(\bm{x})=q^{-n}\sum_{\ell=0}^n \hat{h}_\ell \widehat {1_{S_{\ell}^c}}(\bm{x})\,,
\end{equation}
where $\hat{h}_\ell\geq 0$ and $\hat{h}_0>0$ will be optimized later.
Since $\hat{g}(\bm{\omega})=0$, $\bm{\omega}\in S_\ell\,, \ell>0$, setting $f(\bm{x})=g(\bm{x})h(\bm{x})$ gives
$\hat{f}(\bm{0})=q^{-n}(\hat{g}*\hat{h})(\bm{0})=q^{-n}\hat{g}(\bm{0})\hat{h}_0$. So, the bound \eqref{eq:f0fhat0} becomes 
\begin{equation*}
|\matc|\leq \left(q^n\frac{g(\bm{0})}{\hat{g}(\bm{0})}\right)\left(q^n \frac{h(\bm{0})}{\hat{h}_0}\right)\,.
\end{equation*}
The first term above is precisely Lov\'asz bound and corresponds to the term $\log \theta(C_q)$ in the right hand side of  \eqref{eq:R_LP'_RLP}. We now show that the second term corresponds to the linear programming bound of an imaginary ``Hamming scheme'' with a special non-integer alphabet size $q'=1+\cos(\pi/q)^{-1}$, which is the second term in equation \eqref{eq:R_LP'_RLP}. To do this, define analogously to $S_\ell^c$ the spheres
 \[
S_u^1 = \{\bm{x}:|\{i:x_i=\pm 1 \}|=u,\;  |\{i:x_i=0 \}|=n-u\} \,.
\]
Our constraint is that $h(\bm{x})\leq 0$ if $\bm{x}\in S_u^1$, $u\geq d$. Direct computation shows that for  $\bm{x}\in S_u^1$,
\begin{align*}
\widehat {1_{S_{\ell}^c}}(\bm{x}) & =\sum_{j=0}^\ell \binom{u}{j}\binom{n-u}{\ell-j}(-1)^j2^\ell (\cos(\pi/q))^j\,,\qquad (\bm{x}\in S_u^1)\\
& = (2\cos(\pi/q))^\ell  K_\ell(u;q'),\qquad (q'=1+\cos(\pi/q)^{-1}\,),
\end{align*}
where $K_\ell(u;q')$ is a Krawtchouck polynomial of degree $\ell$ and parameter $q'$ in the variable $u$. We can thus define
\begin{equation}
\Lambda(u)=h(\bm{x})\,, \bm{x}\in S_u^1\,,\qquad \lambda_\ell=q^{-n} (2\cos(\pi/q))^\ell\cdot  \hat{h}_\ell\,,
\label{eq:htoLambda}
\end{equation} 
and write
\begin{equation}
\label{eq:htoLPq'}
q^n\frac{h(\bm{0})}{\hat{h}_0} = \frac{\Lambda(0)}{\lambda_0}\,,
\end{equation}
where the conditions on $h$ can be restated as
\begin{align*}
\Lambda(u) & =\sum_{\ell=0}^n \lambda_\ell K_\ell(u;q')\,,\quad u=0,\ldots,n\,,\\
\lambda_\ell & \geq 0 \,,\quad \ell \geq 0\,,\\
\Lambda(u) & \leq 0\,,\quad u\geq d\,.
\end{align*}
So, the minimization of \eqref{eq:htoLPq'} is reduced to the standard linear programming problem for the Hamming space, though with a non-integer parameter $q'$. Since the construction of the polynomial used in \cite{mceliece-et-al-1977} and \cite{aaltonen1990new} can be applied verbatim for non-integer values of $q'$ (see also \cite{ismail-simeonov-1998} for the position of the roots of $K_\ell(u;q')$), the claimed bound follows.

\end{IEEEproof}

\subsection{Bound via code spectrum}
\label{sec:new_UP_spectrum}

In this section, for the case $\epsilon=1/2$, we improve the bound derived above, following ideas of Kalai-Linial
\cite{kalai-linial-1995} and Litsyn~\cite{KL00}. The main idea is to show that either the minimum distance is smaller
than what was proved in the last section, or there are exponentially many codewords at the minimum distance, which allows us to derive tighter lower bounds on the probability of error.

\begin{theorem}
Let $\epsilon=1/2$ and $q$ be odd. We have the bound
\begin{equation}
E(R) \leq \max_{\delta,\tau}\left[\min\left(\delta \log2, \tau\log 2-\min(R-(\log q-h_3(\tau))\,,\delta/2\log 2 )\right)\right]
\end{equation}
where the maximum is over $\delta\in [h_3^{-1}(\log q-R), s]$, and $\tau\in[\delta, s]$, and where $s$ is such that $R=\log \theta(C_q)+R_{LP}(q',s)$,  having set $q'=q/\theta(C_q)=1+\cos(\pi/q)^{-1}$.
\end{theorem}

\begin{figure}
\centering
\includegraphics[width=0.8\linewidth]{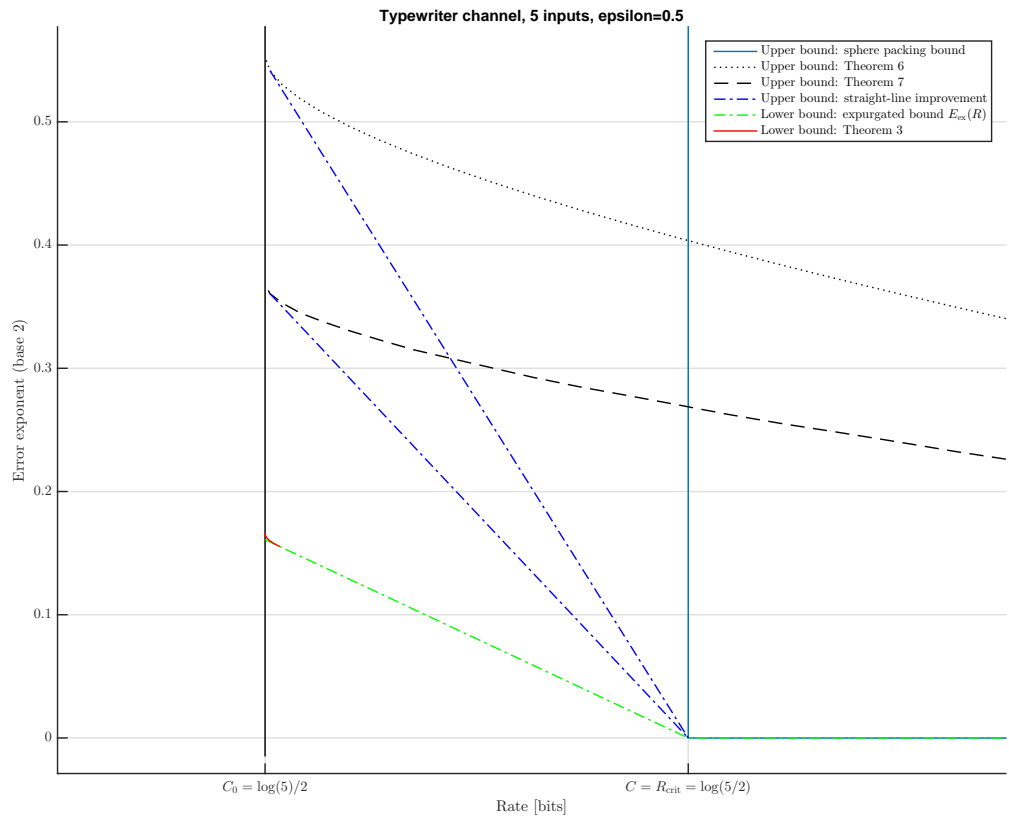}
\caption{Bounds on $E(R)$ for a typewriter channel with 5 inputs, $\epsilon=1/2$.}
\label{fig:q5_KL}
\end{figure}

\begin{IEEEproof}
We first derive a lower bound on the spectrum of the code and then we use it to lower bound the probability of error.
We start again from the equality \eqref{eq:Plancharel} but we now follow the procedure in \cite{kalai-linial-1995}.
As in the previous section, let $\matc$ be a given code of length $n$ with rate $R$ and minimum distance $d=(\delta+o(1))n$. Let again $f$ be a function which is constant on spheres, that is $f(\bm{x})=F(w(\bm{x}))$ such that $\hat{f}\geq 0$ and $\hat{f}(\bm{0})>0$. Now, however, we assume $f(\bm{x})\leq 0$ if $w(\bm{x})\geq (s+o(1)) n$, for some $s\geq \delta$ that we will optimize later. For notational convenience, we treat $\delta n$ and $s n$ as integers neglecting the operations of rounding to integers, which has no practical impact on the asymptotic analysis.  

Set, for $i>\delta n$, 
\begin{equation}
A_i=\frac{1}{|\matc|}|\{(k,j):d(\bm{x}_k,\bm{x}_j)=i\}|
\end{equation}
Upper bounding the left hand side of \eqref{eq:Plancharel} by the sum over positive terms only and lower bounding the right hand size by the zero-frequency term $q^{-n}\hat{f}(\bm{0})|\matc|^2$ we obtain
\begin{equation}
|\matc| \leq q^n {\hat{f}(\bm{0})}^{-1}\left(F(0)+\sum_{i=\delta n}^{sn} A_i F(i)\right)
\label{eq:Kalai-Linial}
\end{equation}
We choose $f$ again of the form $f=gh$ where $g$ is the same as in the previous section and $h$ has the properties expressed in equations \eqref{eq:formofh}, \eqref{eq:htoLambda} and \eqref{eq:htoLPq'} but now (note the use of $s$ in place of $\delta$)
 \begin{align}
\Lambda(u) & =\sum_{\ell=0}^n \lambda_\ell K_\ell(u;q')\,,\quad u=0,\ldots,n\,,\label{eq:LabdaKL}\\
\lambda_\ell & \geq 0 \,,\quad \ell \geq 0\,,\\
\Lambda(u) &  \leq 0 \,,\quad   u\geq (s+o(1))n.
\end{align}
Note that now the sequence $\lambda_\ell$ depends on $s$. We will suppress this dependency in the notation for simplicity, but it will be important to keep it in mind.

Since $g(\bm{x})=\varphi^i$ if $w(\bm{x})=i$, using
\begin{align}
F(i) & = \varphi^i \Lambda(i)\\
\hat{f}(\bm{0}) & =q^{-n}\hat{g}(\bm{0})\hat{h}_0\\
& = \hat{g}(\bm{0})\lambda_0
\end{align}
we get
\begin{align}
|\matc| & \leq \left(q^n\frac{g(\bm{0})}{\hat{g}(\bm{0})}\right) \left(\frac{\Lambda(0)}{\lambda_0}+\sum_{i=\delta n}^{sn} A_i \varphi^i \frac{ \Lambda(i)}{\lambda_0}\right)\\
& = (\theta(C_q))^n\left(\frac{\Lambda(0)}{\lambda_0}+\sum_{i=\delta n}^{sn} A_i  \frac{\varphi^i \Lambda(i)}{\lambda_0}\right)
\end{align}
Taking logarithms and dividing by $n$, 
\begin{equation}
R\leq \log\theta(C_q) + \frac{1}{n}\log \left(\frac{\Lambda(0)}{\lambda_0}+\sum_{i=\delta n}^{sn} A_i \varphi^i \frac{ \Lambda(i)}{\lambda_0}\right)\label{eq:boundR-KL-like}
\end{equation} 
Considering only the dominating term in the parenthesis, we obtain
\begin{equation}
R	 \leq \log \theta(C_q)+ \max\left\{\frac{1}{n}\log \frac{\Lambda(0)}{\lambda_0}, \max \left\{\frac{1}{n}\log \left(A_i  \frac{\varphi^i \Lambda(i)}{\lambda_0}\right), i=\delta n\ldots s n \right\}\right\} (1+o(1))
\end{equation}

The conditions in \eqref{eq:LabdaKL} for $\Lambda$ allow us to employ the MRRW assignment, which gives for the first term in the outer maximum above our previous bound on the rate computed for $s$ in place of $\delta$
\begin{equation}
\frac{1}{n}\log \frac{\Lambda(0)}{\lambda_0}=R_{LP}(q',s)(1+o(1))
\end{equation}
For the inner maximum, the term of index $i=(\tau+o(1))n$ gives a contribution
\begin{equation}
\tau\log(\varphi)+b_\tau+\frac{1}{n}\log\left(\frac{\Lambda(\tau n)}{\lambda_0}\right)+o(1)
\label{eq:threeterms}
\end{equation}
where 
\begin{equation}
b_\tau=\frac{1}{n}\log A_{\tau n}.
\end{equation}
The last term in equation \eqref{eq:threeterms} is asymptotic to (see \cite[Prop. 3.2]{kalai-linial-1995}, of which the following is a $q$-ary extension)
\begin{equation}
2 k(\alpha(s),\tau)-h_{q'}(\alpha(s))
\end{equation}
where $\alpha(s)=h_{q'}^{-1}(R_{LP}(s))$ and, for $a,b\in(0,1)$,  $k(a,b)$ is defined in terms of asymptotic values of the Krawtchouck polynomials as
\begin{equation}
k(a,b)=\lim_{n\to\infty} \frac{1}{n}\log|K_{\lfloor an \rfloor}(\lfloor bn \rfloor;q')|\,.
\end{equation}

So, if we choose $s$ such that $R>\log \theta(C_q)+R_{LP}(q',s)$, there must exists a $\tau$ in the range $\delta\leq \tau \leq s$ such that
\begin{equation}
R\leq \log\theta(C_q) + \tau\log(\varphi)+b_\tau
+2 k(\alpha(s),\tau)-h_{q'}(\alpha(s))\,.
\end{equation}
We can use the $q'$-ary extention of the bound in \cite[Prop. 3.3]{kalai-linial-1995} 
\begin{equation}
2k(a,b)\leq\log(q')+h_{q'}(a)-h_{q'}(b)\,,
\end{equation}
and noticing that $\theta(C_q)=q/q'$ we obtain
\begin{equation}
R\leq \log q + b_\tau+\tau \log(\varphi)-h_{q'}(\tau)
\end{equation}
or, observing $q'-1=2\varphi$, 
\begin{align}
b_\tau & \geq R-\log(q)-\tau\log(\varphi)+h_{q'}(\tau)\\
& = R-\left(\log(q)-h_3(\tau)\right)\,.
\end{align}

So we get to the conclusion that if $s$ satisfies
\begin{equation}
R>\log\theta(C_q)+R_{LP}(q',s)
\end{equation}
then there is a $\tau$ in $[\delta,s]$ such that
\begin{equation}
b_\tau\geq R-\left(\log(q)-h_3(\tau)\right)\label{eq:btau_LB}\,,
\end{equation}
where $\delta$ is the minimum distance of the code. This bound is of course only of interest if the right hand side is non-negative in the whole interval $[\delta,s]$, that is, if
\begin{equation}
\delta \geq h_3^{-1}(\log(q)-R)\,.
\end{equation}
The conclusion that we get is that either 
$$
\delta \leq h_3^{-1}(\log(q)-R)
$$
or there is a $\tau$ in the interval $\delta\leq \tau \leq s$ such that
\begin{equation}
b_\tau\geq R-\left(\log(q)-h_3(\tau)\right)\,.
\end{equation}
This is our bound on the spectrum of the code. We now proceed to derive from this result a lower bound on the probability of error.

Note that there is at least one codeword with at least $e^{n(b_\tau+o(1))}$ neighbors at distance $\tau n$. We can now bound the probability of error for that codeword. We assume all messages are equally likely. Without loss of generality we can consider a randomized decoder which decides uniformly at random among messages compatible with the output, since they all have the same likelihood.
Now, assume a codeword $\bm{x}_1$ is sent which has a neighbor $\bm{x}_2$ at distance $\tau n$. As we mentioned already, from the point of view of two confusable codewords, our channel is like a binary erasure channel and $\bm{x}_1$ will be indistinguishable from $\bm{x}_2$ if all differences are erased. The probability that  the $\tau n$ differences are erased is $2^{-\tau n}$ and this will lead to an error with probability at least $1/2$. So, due to the presence of $\bm{x}_2$ the probability of error when sending $\bm{x}_1$ is at least
\begin{equation}
\Pe\geq 2^{-n(\tau+o(1))}.\label{eq:Pe-oneword}
\end{equation}

We now need to lower bound the probability of error when sending $\bm{x}_1$ due to the presence of $e^{n b_\tau}$ neighbors. Consider a subset of $M$ such neighbors $\bm{x}_2,\bm{x}_3,\ldots\bm{x}_{M+1}$. Let $A_i$ denote the event that the output sequence is compatible with $\bm{x}_i$. Then we can lower bound the probability of error when sending $\bm{x}_1$ as
\begin{align}
\Pe & \geq \frac{1}{2} \PP\left[\cup_{i>1} A_i\right] \\
& \geq  \frac{1}{2} \sum_{i>1} {P[A_i]^2\over \sum_{j>1} P[A_i \cap A_j]}\,,
\label{eq:decaen}
\end{align}
where we have used de Caen's inequality~\cite{de1997lower}. From the previous discussion we know that $P[A_i]=2^{-\tau n}$. For bounding $P[A_i \cap A_j]$, consider the two \emph{distinct} neighbors $\bm{x}_i$ and $\bm{x}_j$ at distance $\tau n$ from $\bm{x}_1$. Note first that if $d(\bm{x}_i,\bm{x}_j)=\infty$ then there is no channel realization which can make $\bm{x}_1$ indistinguishable from both $\bm{x}_i$ and $\bm{x}_j$, so that $P[A_i \cap A_j]=0$. If, instead, $d(\bm{x}_i,\bm{x}_j)$ is finite (but, remind, at least $\delta n$) then with respect to the three sequences the channel is again like a binary erasure channel, and at least $(\tau+\delta/2)n$ erasures are needed to make $\bm{x}_1$ indistinguishable from both $\bm{x}_i$ and $\bm{x}_j$. So, in this case $P[A_i \cap A_j] \leq 2^{-(\tau+\delta/2)n}$. So, from \eqref{eq:decaen} we get
\begin{equation}
\Pe\geq \frac{1}{2} \cdot \frac{M 2^{-2\tau n}}{2^{-\tau n}+(M-1) 2^{-(\tau+\delta/2) n}}\,.
\end{equation}
This quantity is exponentially asymptotic to $M\cdot 2^{-n(\tau+o(1))}$ whenever $M\leq 2^{n\delta/2}$. So, if $e^{n b_\tau}<2^{n\delta/2}$, we can use $M=e^{n b_\tau}$ and
lower bound the probability of error by
\begin{align*} \Pe & \geq e^{-n\tau\log(2)+n b_\tau}\\ & \geq e^{-n(\tau\log(2)-R+(\log q- h_3 (\tau)))} \end{align*}
while if $e^{n b_\tau}\geq 2^{n\delta/2}$ we can take $M=2^{n\delta/2}$ neighbors of $\bm{x}_1$ and bound $\Pe$
as
\begin{equation}\label{eq:decaen_final}
\Pe \geq e^{-n(\tau\log(2)-\min(R-(\log(q)-h_3(\tau))\,,\delta/2\log(2)))}.
\end{equation}

In all, we showed that if $R>\log\theta(C_q)+R_{LP}(q',s)$, then either $\delta \leq h_3^{-1}(R-\log(q))$, in which case 
$$
\Pe \geq e^{-n (h_3^{-1}(R-\log(q))\log(2)+o(1))}\,,
$$
or there is a $\tau\in[\delta,s]$ such that~\eqref{eq:decaen_final} holds.
Of course, we also always have the bound
$$
\Pe \geq e^{-n (\delta\log(2)+o(1))}.
$$

Finally, since we do not know the value of $\delta$, we can consider the most optimistic case and write
\begin{equation}
-\frac{1}{n}\log \Pe \leq \max_{\delta,\tau}\left[\min\left(\delta \log2, \tau\log(2)-\min(R-(\log(q)-h_3(\tau))\,,\delta/2\log(2))\right)\right]\,.
\end{equation}
where the maximum is over $\delta\in [h_3^{-1}(R-\log(q)), s]$ and $\tau\in[\delta, s]$, and $s$ is such that $R=\log \theta(C_q)+R_{LP}(q',s)$. This completes the proof of the theorem.

\end{IEEEproof}

\begin{remark}
\label{rem:spectrum_evenq}
It is worth pointing out that the proof of the theorem could be extended to the case of even $q$, with the only difference that $q'=2$, $\theta(C_q)=q/2$ and $\varphi=1/2$ in that case. However, for even $q$ the assumption $\epsilon=1/2$ is not really interesting since $C_0=C=\log(q/2)$. What is instead interesting is that the bounds derived on the spectrum would mainly differ from the bounds derived  for the binary case in \cite{kalai-linial-1995} for the presence of a coefficient $\varphi^i=2^{-i}$  in the $i$-th term of the summation in the right hand side of equation \eqref{eq:boundR-KL-like} (other than a shift of $\log(q/2)$ on the $R$ axis of course). This agrees with the relation in equation \eqref{eq:fromAtoBspectrum} which expresses the spectrum of our codes for the even length cycle in terms of the spectrum of the used binary code. This should be compared to what we said about Theorem \ref{Th:binary_JPL2}, mentioning that it fails to spot the presence of a high number of neighbors even for even values of $q$. The approach developed in this section suggests that it is indeed possible to prove the presence of those neighbors but unfortunately this would only be useful for the case $\epsilon<1/2$, for which it is difficult to convert efficiently the bound on the spectrum to a bound on the probability of error. We observe that this difficulty arises from the asymmetry introduced by the condition $\epsilon<1/2$ which makes the channel not \emph{pairwise reversible} in the sense of \cite{shannon-gallager-berlekamp-1967-2}. In the case where $C_0=0$, the tightness of the expurgated bound at $R=0$ for such channels was obtained using Berlekamp's complicated procedure \cite{berlekamp-thesis}, \cite{shannon-gallager-berlekamp-1967-2}. Still, that method essentially works to achieve a bound on the minimum distance between two sequences. So, we believe that a very interesting result worth pursuing would be the extension of the bound presented in this section to the case of even $q$ and $\epsilon<1/2$, which might need the combination of Berlekamp's  technique with the Kalai-Linial method and would perhaps allow to close the gap between upper and lower bounds at $R=\log(q/2)$ for even $q$ (see Remark \ref{rem:gap_evenq}).

\end{remark}

\section{Acknowledgments}
The research was supported by the NSF grant CCF-13-18620, by the NSF Center for Science of Information (CSoI) 
under grant agreement CCF-09-39370 and by the Italian Ministry of Education under grant PRIN 2015 D72F16000790001. This work was initiated while the authors were visiting the Simons Institute for the Theory of Computing at UC Berkeley, whose support is gratefully acknowledged.
%

\begin{thebibliography}{10}
\providecommand{\url}[1]{#1}
\csname url@samestyle\endcsname
\providecommand{\newblock}{\relax}
\providecommand{\bibinfo}[2]{#2}
\providecommand{\BIBentrySTDinterwordspacing}{\spaceskip=0pt\relax}
\providecommand{\BIBentryALTinterwordstretchfactor}{4}
\providecommand{\BIBentryALTinterwordspacing}{\spaceskip=\fontdimen2\font plus
\BIBentryALTinterwordstretchfactor\fontdimen3\font minus
  \fontdimen4\font\relax}
\providecommand{\BIBforeignlanguage}[2]{{%
\expandafter\ifx\csname l@#1\endcsname\relax
\typeout{** WARNING: IEEEtran.bst: No hyphenation pattern has been}%
\typeout{** loaded for the language `#1'. Using the pattern for}%
\typeout{** the default language instead.}%
\else
\language=\csname l@#1\endcsname
\fi
#2}}
\providecommand{\BIBdecl}{\relax}
\BIBdecl

\bibitem{dalai-polyanskiy-2016}
M.~Dalai and Y.~Polyanskiy, ``{B}ounds on the {R}eliability of a {T}ypewriter
  {C}hannel,'' in \emph{Proc. IEEE Intern. Symp. Inform. Theory}, 2016, pp.
  1715--1719.

\bibitem{shannon-gallager-berlekamp-1967-1}
C.~E. Shannon, R.~G. Gallager, and E.~R. Berlekamp, ``{L}ower {B}ounds to
  {E}rror {P}robability for {C}oding in {D}iscrete {M}emoryless {C}hannels.
  {I},'' \emph{Information and Control}, vol.~10, pp. 65--103, 1967.

\bibitem{gallager-book}
R.~G. Gallager, \emph{Information Theory and Reliable Communication}.\hskip 1em
  plus 0.5em minus 0.4em\relax Wiley, New York, 1968.

\bibitem{shannon-1956}
C.~E. Shannon, ``{T}he {Z}ero-{E}rror {C}apacity of a {N}oisy {C}hannel,''
  \emph{IRE Trans. Inform. Theory}, vol. IT-2, pp. 8--19, 1956.

\bibitem{lovasz-1979}
L.~Lov{\'a}sz, ``{O}n the {S}hannon {C}apacity of a {G}raph,'' \emph{IEEE
  Trans. Inform. Theory}, vol.~25, no.~1, pp. 1--7, 1979.

\bibitem{baumert-et-al-1971}
L.~D. Baumert, R.~J. McEliece, E.~Rodemich, R.~H.~C. Jr., R.~Stanley, and
  H.~Taylor, ``{A} {C}ombinatorial {P}acking {P}roblem,'' \emph{Proc. SIAM
  AMS}, vol.~4, pp. 97--108, 1971.

\bibitem{bohman-2003-I}
T.~Bohman, ``A limit theorem for the shannon capacities of odd cycles i,''
  \emph{Proc. Am. Math. Soc.}, vol. 131, pp. 3559--3569, 2003.

\bibitem{mathew-ostergard-2016}
K.~A. Mathew and P.~R.~J. {\"O}sterg{\aa}rd, ``New lower bounds for the shannon
  capacity of odd cycles,'' \emph{Designs, Codes and Cryptography}, pp. 1--10,
  2016.

\bibitem{katsman-tsfasman-vladut-1992}
G.~L. Katsman, M.~A. Tsfasman, and S.~G. Vl\u{a}du\c{}t,
  ``\BIBforeignlanguage{English}{Spectra of linear codes and error probability
  of decoding},'' in \emph{\BIBforeignlanguage{English}{Coding Theory and
  Algebraic Geometry}}, ser. Lecture Notes in Mathematics, 1992, vol. 1518, pp.
  82--98.

\bibitem{delsarte-1973}
P.~Delsarte, ``An {A}lgebraic {A}pproach to the {A}ssociation {S}chemes of
  {C}oding {T}heory,'' \emph{Philips Res. Rep.}, vol.~10, 1973.

\bibitem{mceliece-et-al-1977}
R.~McEliece, E.~Rodemich, H.~Rumsey, and L.~Welch, ``New upper bounds on the
  rate of a code via the {D}elsarte-{M}ac{W}illiams inequalities,''
  \emph{Information Theory, IEEE Transactions on}, vol.~23, no.~2, pp. 157 --
  166, mar 1977.

\bibitem{litsyn-1999}
S.~Litsyn, ``New upper bounds on error exponents,'' \emph{IEEE Transactions on
  Information Theory}, vol.~45, no.~2, pp. 385--398, Mar 1999.

\bibitem{barg-mcgregor-2005}
A.~Barg and A.~McGregor, ``Distance distribution of binary codes and the error
  probability of decoding,'' \emph{Information Theory, IEEE Transactions on},
  vol.~51, no.~12, pp. 4237 -- 4246, dec. 2005.

\bibitem{kalai-linial-1995}
G.~Kalai and N.~Linial, ``On the distance distribution of codes,'' \emph{IEEE
  Trans. Inform. Theory}, vol.~41, no.~5, pp. 1467--1472, Sep 1995.

\bibitem{gallager-1965}
R.~G. Gallager, ``A {S}imple {D}erivation of the {C}oding {T}heorem and {S}ome
  {A}pplications,'' \emph{IEEE Trans. Inform. Theory}, vol. IT-11, pp. 3--18,
  1965.

\bibitem{jelinek-1968}
F.~Jelinek, ``{E}valuation of {E}xpurgated {E}rror {B}ounds,'' \emph{IEEE
  Trans. Inform. Theory}, vol. IT-14, pp. 501--505, 1968.

\bibitem{dalai-TIT-2013}
M.~Dalai, ``{L}ower {B}ounds on the {P}robability of {E}rror for {C}lassical
  and {C}lassical-{Q}uantum {C}hannels,'' \emph{IEEE Trans. Inform. Theory},
  vol.~59, no.~12, pp. 8027 -- 8056, 2013.

\bibitem{korn-1968}
I.~Korn, ``On the {L}ower {B}ound of {Z}ero-{E}rror {C}apacity,'' \emph{IEEE
  Trans. on Inform. Theory}, vol.~14, no.~3, pp. 509 -- 510, may 1968.

\bibitem{motzkin1965maxima}
T.~S. Motzkin and E.~G. Straus, ``Maxima for graphs and a new proof of a
  theorem of tur{\'a}n,'' \emph{Canad. J. Math}, vol.~17, no.~4, pp. 533--540,
  1965.

\bibitem{mceliece-et-al-1978}
R.~J. Mc{E}liece, E.~Rodemich, and H.~Rumsey, ``The {L}ov{\'a}sz bound and some
  generalizations,'' \emph{J. Combin. Inform. System Sci.}, vol.~3, pp.
  134--152, 1978.

\bibitem{poltyrev-1994}
G.~Poltyrev, ``Bounds on the decoding error probability of binary linear codes
  via their spectra,'' \emph{IEEE Transactions on Information Theory}, vol.~40,
  no.~4, pp. 1284--1292, Jul 1994.

\bibitem{barg-forney-2002}
A.~Barg and G.~D. Forney, ``Random codes: minimum distances and error
  exponents,'' \emph{IEEE Transactions on Information Theory}, vol.~48, no.~9,
  pp. 2568--2573, Sep 2002.

\bibitem{berlekamp-thesis}
E.~R. Berlekamp, ``Block coding with noiseless feedback,'' Ph.D. dissertation,
  MIT, Cambridge, MA, 1964.

\bibitem{shannon-gallager-berlekamp-1967-2}
C.~E. Shannon, R.~G. Gallager, and E.~R. Berlekamp, ``{L}ower {B}ounds to
  {E}rror {P}robability for {C}oding in {D}iscrete {M}emoryless {C}hannels.
  {II},'' \emph{Information and Control}, vol.~10, pp. 522--552, 1967.

\bibitem{csiszar-korner-bookv2}
I.~Csisz{\'{a}}r and J.~K{\"{o}}rner, \emph{Information Theory: Coding Theorems
  for Discrete Memoryless Systems}.\hskip 1em plus 0.5em minus 0.4em\relax
  Cambridge University Press, 2011.

\bibitem{dalai-polyanskiy-2015a}
M.~Dalai and Y.~Polyanskiy, ``Bounds for codes on pentagon and other cycles,''
  \emph{arxiv:1508.03020}, Aug. 2015.

\bibitem{cullina-dalai-polyanskiy-2016}
D.~Cullina, M.~Dalai, and Y.~Polyanskiy, ``{R}ate-distance tradeoff for codes
  above graph capacity,'' in \emph{Proc. IEEE Intern. Symp. Inform. Theory},
  2016, pp. 1331--1335.

\bibitem{schrijver-1979}
A.~Schrijver, ``A comparison of the {D}elsarte and {L}ov{\'a}sz bounds,''
  \emph{IEEE Trans. on Inform. Theory}, vol.~25, no.~4, pp. 425 -- 429, jul
  1979.

\bibitem{aaltonen1990new}
M.~Aaltonen, ``A new upper bound on nonbinary block codes,'' \emph{Discrete
  Mathematics}, vol.~83, no.~2, pp. 139--160, 1990.

\bibitem{ismail-simeonov-1998}
M.~E.~H. Ismail and P.~Simeonov, ``Strong {A}symptotics for {K}rawtchouk
  {P}olynomials,'' \emph{J. Comput. Appl. Math.}, vol. 100, no.~2, pp.
  121--144, Dec. 1998.

\bibitem{KL00}
O.~Keren and S.~Litsyn, ``A lower bound on the probability of decoding error
  over a {BSC} channel,'' in \emph{Electrical and ELectronic Engineers in
  Israel. The 21st IEEE Convention of the}, 2000, pp. 271--273.

\bibitem{de1997lower}
D.~De~Caen, ``A lower bound on the probability of a union,'' \emph{Discrete
  mathematics}, vol. 169, no. 1-3, pp. 217--220, 1997.

\end{thebibliography}

\end{document}